\documentclass[aps,prb, twocolumn,amsmath,amssymb, superscriptaddress,citeautoscript,floatfix]{revtex4-2}

\usepackage[utf8]{inputenc}
\usepackage[T1]{fontenc}
\usepackage{braket}
\usepackage{stix}
\usepackage{amsmath}
\usepackage{placeins}
\usepackage{graphicx}
\usepackage{comment}
\usepackage{hyperref}
\usepackage{xcolor}
\usepackage{bm}
\usepackage{tikz}
\usepackage{changepage} 
\usepackage{amssymb}
\usepackage{indentfirst}
\usepackage{adjustbox}
\usepackage{pgffor}
\usepackage[percent]{overpic}
\usepackage{float}
\usepackage{setspace}
\usepackage{titlesec}

\allowdisplaybreaks

\hypersetup{
    colorlinks=true,      
    linkcolor=black,      
    filecolor=black,        
    urlcolor=blue,         
    citecolor=black,       
    pdfborder={0 0 0},     
}
\begin{document}
\title{Chirality reversal at finite magnetic impurity strength  and 
local  signatures of a topological phase transition}

\author{Ruiqi Xu}
\affiliation{School of Physics, Georgia Institute of Technology, Atlanta, GA 30332, USA}
\author{Arnab Seth}
\affiliation{School of Physics, Georgia Institute of Technology, Atlanta, GA 30332, USA}
\author{Itamar Kimchi}
\affiliation{School of Physics, Georgia Institute of Technology, Atlanta, GA 30332, USA}

\date{December 9, 2025}

\begin{abstract}
    We study the honeycomb lattice with a single magnetic impurity modeled by adding imaginary next-nearest-neighbor hopping $ih$ on a single hexagon. %
    This Haldane defect gives a topological mass term to the gapless Dirac cones and generates chirality. For a small density of defects Neehus et al.\ [PRL 135, 126604 (2025)] found that the system's chirality reverses at a critical $h_c\approx 0.95$ associated with an unexpected tri-critical point of Dirac fermions at zero defect density. We investigate this zero-density limit by analyzing a single defect and computing
    two experimentally relevant measures of chirality: (1) orbital magnetization via local Chern marker, a bulk probe of all occupied states; and (2) electronic currents of low-energy states. %
    Both probes
    show a chirality reversal at a critical $h_c\approx0.9-1.0$. Motivated by this consistency
    we propose a defect-scale  toy model 
    whose low energy states
    reverse their chirality at $h'_c%
    \approx 0.87$.
    Remarkably, the same pair of zero energy bound states also generate the critical point $h_c$ in the  
    full %
    impurity projected T-matrix.
    Our results show how the  chirality reversal produced by an impurity can be observed either in local probes or in the global topology  
    and suggest a possible role of the microscopic defect structure at the critical point.
\end{abstract}
\maketitle

\section{Introduction}
\begin{figure*}[ht]
  \centering
  \begin{adjustbox}{valign=t}
    \begin{overpic}[width=0.24\textwidth]{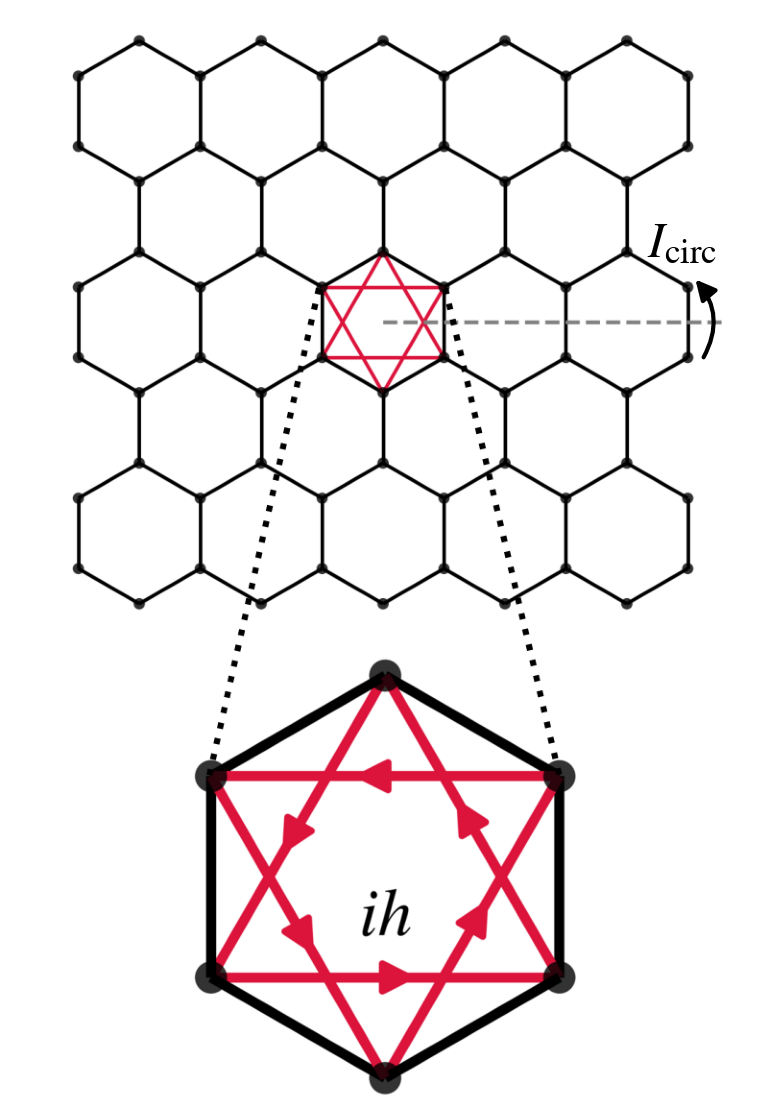}
      \put(3,100){\small(a)}
      \put(6,-8){\text{Haldane hexagon defect}}
    \end{overpic}
  \end{adjustbox}%
  \hfill
  \begin{adjustbox}{valign=t}
    \begin{overpic}[width=0.75\textwidth]{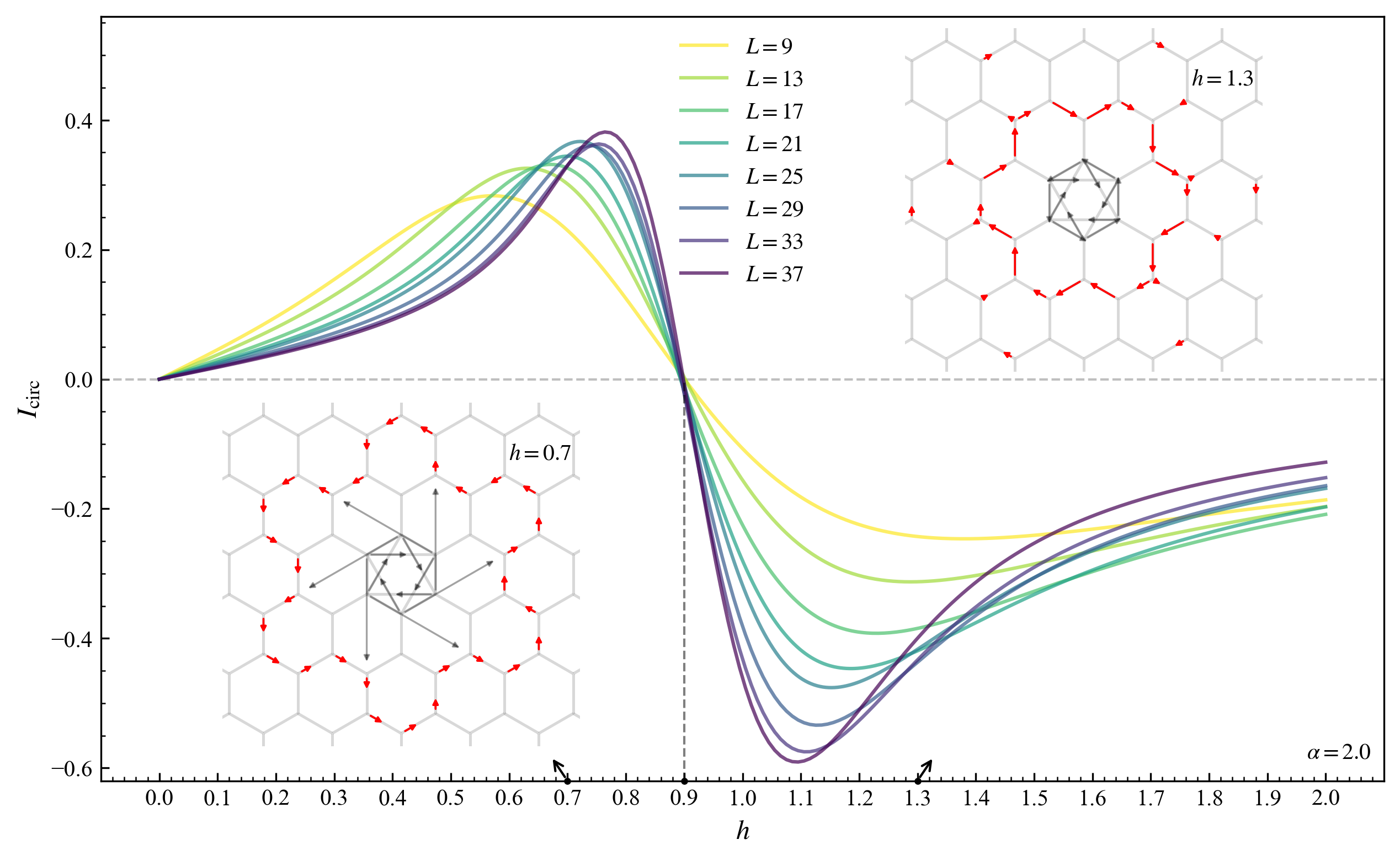}
      \put(3,61){\small(b)}
    \end{overpic}
  \end{adjustbox}

  \caption{Chirality reversal transition observable as a sign change of  current circulation around a Haldane hexagon defect. (a) The model. Honeycomb lattice with a single Haldane hexagon defect consisting of imaginary hopping $i h$ (oriented counterclockwise, red arrows) on second-neighbor bonds.   
 Circulating current $I_{\text{circ}}$ is defined as the total current flow across the horizontal grey dashed line $+\hat{x}$. 
  (b) Sign flip of $I_{\text{circ}}$ at the  critical point $h=h_c$. 
  Main panel: $I_{\text{circ}}$ as a function of impurity scattering strength $h$ showing a sign flip of $I_{\text{circ}}$ at $h_c \approx 0.9$ independent of system size $L$. Current is computed from low energy states (Gaussian energy filter width  $\sigma=\alpha/L$ with $\alpha=2.0$).
  Inset: spatial distribution of currents for $L=25$ at $h$ below and above the transition ($0.7$ and $1.3$; see Appendix~\ref{appendix:visualization} for additional values). 
    While 
    currents on most lattice bonds (red arrows) clearly show  chirality reversal,
    currents on the core defect bonds (thin black arrows) show more complicated behavior which is analyzed further below.}
  \label{fig1}
\end{figure*}

The quantum Hall effect with its topologically protected chiral edge modes has analogues in electronic insulators with Chern bands as exemplified by Haldane's honeycomb model \cite{haldane_model_1988}. 
The model begins with spinless fermions hopping on the honeycomb (graphene) lattice with chemical potential tuned to the two 2D Dirac cones. Time reversal breaking (TRB) is then incorporated by adding imaginary  hopping on next-nearest-neighbor (NNN) bonds of the honeycomb lattice.
Microscopically the imaginary hopping term is quite general and could describe orbital magnetic fields or arise via superexchange in the presence of any magnetic background.
The interesting effects of the TRB term can be seen by considering its action on the gapless Dirac cones: it  opens a topological gap and generates a Chern number set by the sign of the imaginary hopping. As in the Quantum Hall effect, the bulk band with nonzero Chern number has a corresponding chiral edge state whose circulating current reflects the bulk topology.

What happens to the electronic structure if time reversal is broken only by a dilute density of magnetic impurities? 
Extensive earlier studies have discussed the generation of spin magnetism, such as the induction of Stoner magnetism in graphene via various local defects \cite{wehling_local_2007, yazyev_defectinduced_2007, ugeda_missing_2010} or via  boundary effects in a finite sample \cite{lee_magnetic_2005}.
However the effects of defects on orbital magnetization, arising from the electronic bands even in the absence of spin magnetization, have not been as well explored. The spinless Haldane model offers an avenue for accessing this limit. 
Each magnetic scatterer can be modeled as a defect hexagon.  The imaginary NNN bonds are then added only within these isolated defect hexagons (Fig~\ref{fig1}). Such Haldane hexagon defects provide the effective description of a magnetic impurity or any other TRB scattering, with no spin degree of freedom explicitly involved.

The resulting noninteracting spinless Hamiltonian,  including the magnetic impurities modeled as Haldane hexagon ``$\hexagon'$'' defects is $H=H_{0}+ \sum_{\hexagon'} V_{\hexagon'}$, with

\begin{align}     \label{eq1}
H_{0}= -\sum_{\langle jj'\rangle} c^\dagger_{j} c_{j'} , \quad  V_{\hexagon'}= i h \sum_{\langle\langle jj''\rangle\rangle\in \hexagon'} c^\dagger_{j} c_{j''}
\end{align}
where the addition of the Hermitian conjugate is implied. 
Here we use $H_{0}$ consisting of nearest neighbor hopping on the honeycomb (graphene) lattice, with unit magnitude, but any symmetry preserving hopping will preserve the Dirac cones \cite{kimchi_featureless_2013} and is expected to produce similar behavior.
The $\hexagon'$ denotes the defect hexagon which gains the TRB imaginary hopping $ih$ with $h$ giving the TRB impurity strength.  Since imaginary hopping is directional, sites $j$ and $j''$ are chosen in counterclockwise order around this hexagon center  (Fig~\ref{fig1}(a)). 
The Haldane honeycomb model serves as an opposite limit of this Hamiltonian and corresponds to the case where $V_{\hexagon'}$ appears at every hexagon, producing a Chern number $C= \text{sgn}[h]$ independent of the magnitude of $h$.

The Hamiltonian Eqn.~\ref{eq1} was recently studied by Neehus, Pollmann and Knolle \cite{neehus_genuine_2025}.
By numerically computing the Hall conductivity 
and disorder averaged Bott index   they found that the chirality generated by the defects is reversed above a critical value of $h_c$ for finite impurity densities. This chirality reversal critical point appears to be a tri-critical point of the model at zero impurity density, 
connecting a finite impurity density thermal metal phase to the two chirality reversed topological phases. 
Defects have been known to modify local chirality in other settings \cite{jha_impurityinduced_2017, karmakar_disorderinduced_2021, gonzalez_impurity_2012, michel_bound_2024} including by extending topological phases into Anderson insulators \cite{li_topological_2009,groth_theory_2009,meier_observation_2018}
but this model's topological phase with reversed chirality  relies on the presence of (dilute) impurities and thus appears to be distinct \cite{neehus_genuine_2025}. 
The $h_c$ chirality reversal tri-critical point's appearance at zero impurity density for this Dirac cone system (particle-hole symmetry class D \cite{ryu_topological_2010, kitaev_periodic_2009}) 
is an unexpected result \cite{ludwig_integer_1994,cho_criticality_1997,bocquet_disordered_2000,merz_twodimensional_2002,gruzberg_randombond_2001,chalker_thermal_2001,medvedyeva_effective_2010,mildenberger_griffiths_2006}
which is not sufficiently understood.

This surprising critical point arising from  dilute impurities motivated us to investigate the emergence of chirality reversal for a single isolated defect. %
The single defect serves as a building block for the low impurity density limit and allows us to analyze the spatial dependence of chirality probes.
Consider the measure of chirality consisting of electronic currents. 
Recall that for gapped topological states with nonzero Chern number the bulk-boundary correspondence implies that the topology can be observed on the system's boundary through chiral low energy states, albeit sometimes in  a manner complicated by the lattice  \cite{huang_vanishing_2014}. %
Numerically computing probability currents for this model on finite open systems we find that the average current circulation is reversed  at a finite $h_c\approx 0.9$ (Fig.~\ref{fig1}(b)) corresponding to the critical point of the Chern number flip at $h_c\approx 0.95$  (Fig.~\ref{fig2}). 
To further investigate this behavior we perform numerical computations of the local Chern marker \cite{bianco_mapping_2011, bianco_orbital_2013, ceresoli_orbital_2006} (measurable via orbital magnetization) as well as analytic computations of the impurity projected T-matrix.
We analyze the results in terms of local defect-scale probes of the chirality and in terms of a defect-scale toy model we introduce. Surprisingly, this toy model captures a chirality reversal at a parameter value which is closely related to the true global $h_c$ (tri-)critical point. 
Our results relate experimental probes at different length scales, and  suggest 
that though the chirality reversal topological phase transition of this model requires the impurity density to be small (and nonzero), it nevertheless 
involves local features at the lattice scale of the defect core.

The rest of this manuscript is structured as follows.
We begin in Sec.~II with a detour to the case of defect superlattices  %
which enable a connection to the conventional Chern number defined for periodic systems.
We show how the Dirac cone projected T-matrix, as defined by Ref.~\cite{neehus_genuine_2025},  explains that defect supercell size determines whether the chirality is reversed or is extinguished at the (supercell-independent) phase transition. 
This analysis also explains the necessary role played by the open boundaries in our subsequent numerical work. In Sec.~III we report our numerical results for the circulating probability currents and the local Chern marker (proportional to electronic orbital magnetization) and discuss their similarities and differences in identifying the chirality reversal phase transition. 
In Sec.~IV we analyze the spatial distribution of the probability currents and use the results to motivate a toy model of the bound state wavefunction phase windings at the defect core which shows chirality reversal at a critical $h_c'$. 
We then compute the impurity projected T-matrix to identify the two eigenstates responsible for the critical point $h_c$, finding that these are identical to the states producing $h_c'$ in the toy model. In Sec.~V we discuss implications of these results. 
Further details are given in Appendices.

\section{Defect superlattices and Dirac cone projected T-matrix}
\label{sec:T-matrix}

\begin{figure*}[ht]
  \centering
  \begin{adjustbox}{valign=t}
    \begin{overpic}[width=0.48\textwidth]{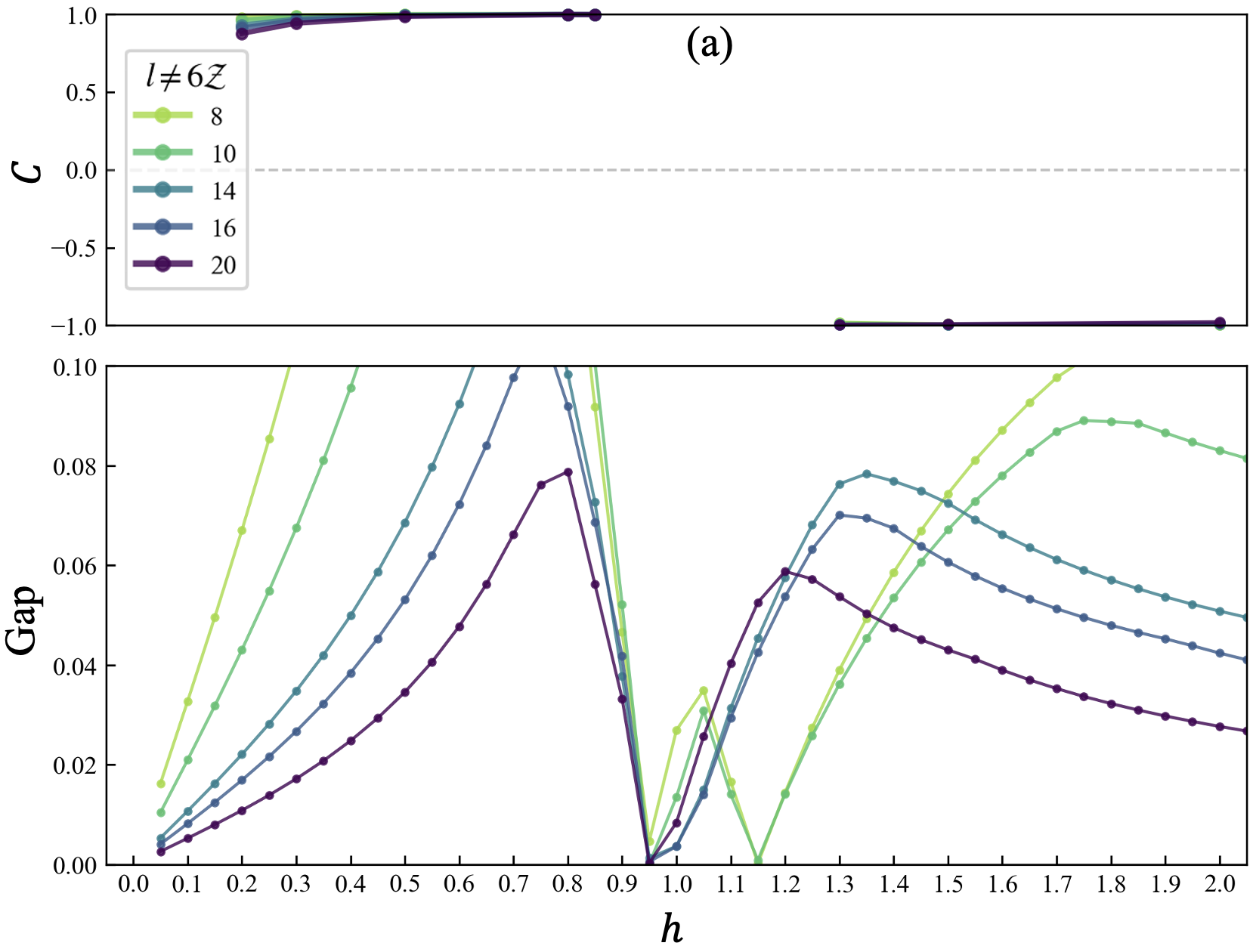}
    \end{overpic}
  \end{adjustbox}%
  \hfill
  \begin{adjustbox}{valign=t}
    \begin{overpic}[width=0.48\textwidth]{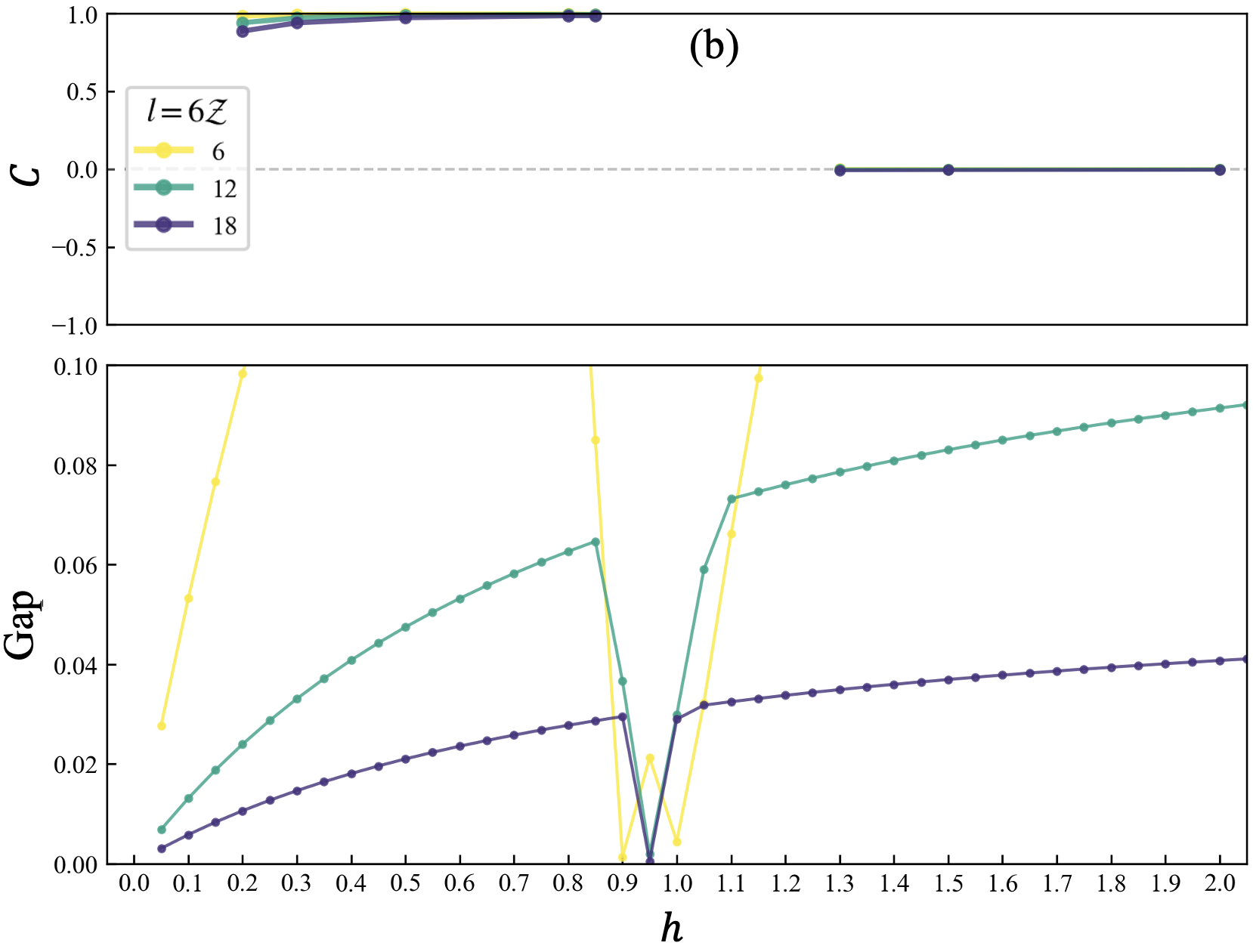}
    \end{overpic}
  \end{adjustbox}
  \caption{Chirality transitions for $l \times l$ defect superlattices.   Top:  Chern number topological invariant showing a transition as a function of $h$ near $h_c\approx0.95$. 
  Bottom: the energy gap also shows the transition. %
  (a) Left: $l\neq6\mathcal{Z}$ arrays ($\mathcal{Z}$ integer) with distinct $K$, $K'$ Dirac cones show the chirality reversal transition. (b) Right: $l=6\mathcal{Z}$ arrays with Dirac cones overlapping at $\Gamma$ show a chirality extinction. The chirality reversal is the generic case. 
  }
    \label{fig2}
\end{figure*}

Before describing the numerical results for a single defect we begin with the naive approach of a superlattice defect array. This setup enables the use of translational symmetry to define Chern number, and serves as a practical starting point for analysis. 

We use a superlattice array whose supercell has $l^2$ sites (with $l\in 2\mathcal{Z}$ with  $\mathcal{Z}$ integers) and place one Haldane hexagon defect in the cell. 
Fig.~\ref{fig2} shows the Chern number and band gap we computed for such periodic arrangement of defects. It is evident that there is a clear distinction between non-$6\mathcal{Z}$ and $6\mathcal{Z}$ cases for $l$: the former undergoes a chirality reversal, while the latter results in chirality extinction, both occurring near the gap closing point at $h_c\approx0.95$.

This behavior and related effects \cite{jiang_stabilizing_2012} can be explained using the T-matrix which is a good approximation for dilute (low density) defects. %
For a single defect problem, the Green's function can be expressed exactly in terms of the T-matrix, which accounts for multiple scatterings off the same defect to all orders. 
\begin{align}
    T_{\hexagon'}(r_0,E)=V_{\hexagon'}\left(1-G(E)V_{\hexagon'}\right)^{-1}
\end{align}
where $r_0$ denotes the location of the defect hexagon $\hexagon'$, and $G(E)$ denotes the Green's function of graphene at energy $E$. Since we are primarily interested in the low energy states (where the topological properties of the system manifest), the T-matrix can be projected to the Dirac cones - particularly considering that the Chern number in Haldane model is determined by the sign of TR broken mass term that opens the gap.
As shown by Ref.~\cite{neehus_genuine_2025}  the Dirac cone projected T-matrix \cite{kot_band_2020} at $E=0$  gives
\begin{align}
\nonumber
    &T_{\hexagon'}\left({{r}_0} , E=0\right)=g(h)\left(2\sqrt{3}\pi\tau^z\sigma^z+ \vec{m}_\text{triv}\cdot
    (\tau^x,\tau^y)
    \sigma^x\right)
\label{eq:T-matrix}
\\ \nonumber
    &g(h)=\frac{18\pi h }{-12\pi^2+\left(3\sqrt{3}+2\pi\right)^2h^2}\\
    &\vec{m}_\text{triv}=h\ (3\sqrt{3}+2\pi) \left(\text{Re},~\text{Im}\right)\left[e^{i({ K-K'})\cdot { r_0}}\right]
\end{align}
with $K=-K'$ denoting the two Dirac cones. All the terms in the T-matrix anticommute with the kinetic Dirac matrices and therefore act as mass terms for Dirac cones. The first term $\tau^z\sigma^z$ breaks TR symmetry and generates chirality (producing Chern number $C=1$ for  $-\tau^z\sigma^z$, resulting in a counterclockwise probability current at the system boundaries). The second mass term preserves TR symmetry and is topologically trivial.

The position dependent phase factor $e^{i({ K-K'})\cdot { r_0}}$ in the trivial mass term is our interest here. For the disordered problem, it cancels out upon summing over the positions $r_0$ of randomly placed defects:
\begin{align}
    \sum_{r_0\in \text{random}} \vec{m}_\text{triv}\propto\sum_{r_0\in \text{random}} e^{i(K-K')\cdot r_0}=0
\end{align} 
Therefore, the averaged T-matrix (or self-energy) reduces to its TR broken term and thus determines the system's chirality. As discussed in Ref.~\cite{neehus_genuine_2025} the denominator of the function $g(h)$ is negative when $h<|h_c|=2\sqrt{3}\pi/(3\sqrt{3}+2\pi)
\approx0.948$, and becomes positive beyond this pole, consistent with the chirality reversal observed numerically near 0.95 at small defect densities.
The sign change in the denominator is a resonance, enabling the topological mass term to change sign at $h_c$ by diverging without crossing zero continuously.

Returning to our array results, the observations can now be readily understood. For $l=6\mathcal{Z}$ supercell sizes, both Dirac cones are folded to the $\Gamma$ point, $K, K'\rightarrow\Gamma$, hence the phase coherently adds up giving rise to a competition between trivial and nontrivial mass terms. 
The $g(h)$ pole at $h_c$ also coincides with the critical $h_c$ where the amplitude of the trivial mass term for a single $r_0$ overcomes the topological mass term. 
Thus for $l=6\mathcal{Z}$ and $h>h_c$, the trivial mass term wins. 
On the other hand, for $l\neq6\mathcal{Z}$, $K,K'=\left(\pm4\pi/3\sqrt{3}l,0\right)$, and the phase cancels out when summed over all the integer $m_x$ similarly to the random disorder case. This provides a  dependence on defect supercell size,
\begin{align}
    \sum_{r_0\in m_x,m_y} \vec{m}_\text{triv}\propto\sum_{m_x,m_y} e^{i(K-K')\cdot r_0}=\left\{\begin{array}{l}
         N_\text{defects} ~~\text{for} ~~l= 6\mathcal{Z} \\
         0  ~~\text{for} ~~l\neq 6\mathcal{Z} 
    \end{array}\right.
\end{align}
This indicates that for $l=6\mathcal{Z}$, the trivial mass wins over the TR broken mass for $h>h_c$ leading to chirality extinction, whereas for $l\neq6\mathcal{Z}$ the TR broken mass dominates for all values of $h$, leading to chirality reversal for $h>h_c$, consistent with our numerical result. 

For a single defect %
the phase factor in trivial mass term remains finite. Thus, the T-matrix approximation predicts chirality extinction instead of chirality reversal: there is chirality for $h<h_c$, which vanishes as $h$ exceeds $h_c$ because the magnitude of trivial mass term becomes larger than the nontrivial one. 

To circumvent this issue for studying a single defect we consider a finite system with open boundaries. We expect the boundaries to provide additional scattering processes beyond those captured by the single defect T-matrix approach, thus eliminating the trivial mass phase factor and re-enabling the chirality reversal transition.
Such boundary scattering can be studied analogously to Ref.~\cite{settnes_patched_2015} to compute how the boundary modifies the various wavefunctions in the system. Since any amount of scattering, or equivalently any amount of wavefunction phase scrambling,  is sufficient to destroy the trivial mass phase factor, we expect the details of the boundary to be unimportant in enabling a single defect chirality reversal.
This expectation is indeed confirmed, as we shall now see.

\section{Two probes of chirality reversal}
\label{sec:OBC}
\begin{figure*}[ht]
    \centering
    \includegraphics[width=0.8\textwidth]{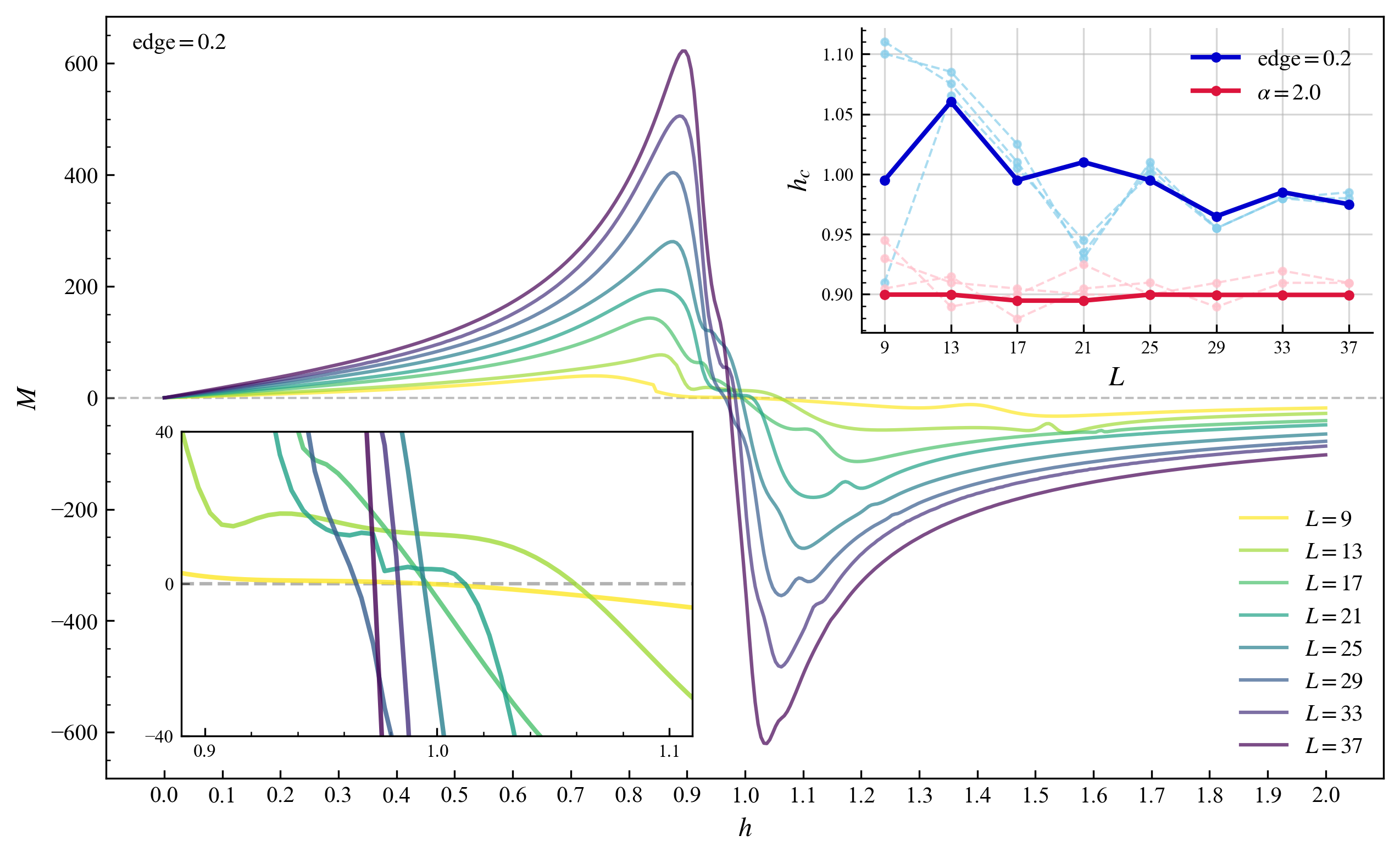}
    \caption{Reversal of orbital magnetization as measured by the integrated local marker $M$. Main panel: $M$ as a function of $h$ over different system sizes. $\text{edge}=0.2$ indicates that boundary strips of width 20\% are taken in both directions, and the sites belonging to the edge region are excluded from $M$ integration. 
    Bottom-left: A zoomed-in view around the chirality reversal transition point $h_c$.
    Top-right: Comparison of $h_c$ obtained from the two different probes:  from current flow $I_{\text{circ}}$ (upper/blue lines) and from integrated local Chern marker $M$ (lower/red lines). Dark blue and dark red solid lines correspond to the main parameter values used ($\alpha=2$ and $\text{edge}=0.2$). The surrounding dashed lines are results under other parameter choices ($\alpha=0.5,1,4$; $\text{edge}=0.3,0.4,0.5$).  The bulk probe $M$ gives slightly higher values of $h_c$ and shows larger finite size effects compared to $I_{\text{circ}}$. }
    \label{fig3}
\end{figure*}

In this section we investigate a model with a single Haldane hexagon defect under open boundary conditions (OBC).  The open boundaries enable us to  use the local Chern marker and chiral probability currents, as will be discussed below. We use these two complementary methods to investigate the system's chirality and find that the system undergoes a chirality reversal transition.

We introduce a Haldane-type defect in the hexagon in the middle of the system by adding six imaginary next-nearest-neighbor (NNN) hopping terms  $h$. System sizes are chosen to preserve inversion symmetry across the defect midpoint.  Fig.~\ref{fig1}(a) shows a schematic for $L=5$, representing a maximum length and width of 5 hexagons. This configuration has approximately $L^2$ hexagons and $2L^2$ sites. 
All model construction and numerical calculations are performed based on the Kwant library \cite{groth_kwant_2014}, a Python package for quantum transport simulations.

\subsection{Orbital magnetization and local Chern marker}

Although the Chern number topological invariant was originally formulated as a momentum space integral, it can also be expressed \cite{bianco_mapping_2011} as the imaginary part of a real space trace of the real-space operator $PXPYP$. Here $X$ and $Y$ are position operators, and $P$ is the occupied-states projector. However, if the real space trace is taken over the entire sample, or in PBC, the result always vanishes, since  the trace of a commutator in finite dimensions is zero.
The local Cherm marker typically requires a finite open system to be well defined \cite{bianco_mapping_2011, bianco_orbital_2013, ceresoli_orbital_2006} (though extensions to PBC in a $k=0$ large supercell limit are being proposed \cite{bau_local_2024}).  
Therefore, to characterize the system's chirality, we compute the sum of local Chern marker $M(r)$ over a bulk region that avoids an ``edge'' region near the boundary:
\begin{equation}
    M = \sum_{\mathbf{r}\in\text{bulk}}M(\mathbf{r})
    =-\frac{4\pi}{A_c}\sum_{\mathbf{r}\in\text{bulk}}\operatorname{Im}\bra{\mathbf{r}}PXPYP\ket{\mathbf{r}}
    \label{eq:chernmarker}
\end{equation}
where the area per unit cell $A_c=\sqrt{3}/2$, in units where Bravais lattice basis vectors have length 1, is here included to give a 2D magnetization as a magnetic moment per area. 
The local Marker $M(r)$ shows one sign in the bulk of the system and a large opposite contribution on sites near the edge of the system, canceling the bulk and resulting in a vanishing total trace. Plots of $M(r)$ along 1D spatial cuts can be found in Appendix~\ref{appendix:marker}.
To compute the integrated $M$, we  define an ``edge'' parameter as the linear fraction of lattice sites to be excluded from the bulk integration window. 
Note that in homogeneous gapped systems the edge region can in principle be chosen to be larger than the correlation length. In contrast, the present case implies a possible dependence on the definition of the edge region due to the extended gapless background. 
In this work we usually set the linear edge region to the smallest value suggested by the $M(r)$ plots, 0.2 ($20\%$), implying that approximately 36\% of sites in the system are excluded from the bulk integration window. 
However we also consider larger definitions of the edge region and find that the overall behavior remains unchanged (Fig. \ref{fig3}), suggesting that the bulk integrated local marker $M$ is robust.

The bulk integrated marker $M$ is plotted in the main panel of Fig. \ref{fig3} as a function of $h$ for different system sizes. A sign flip is observed near $h_c \approx 1$  (see the lower left inset for details), indicating a reversal in the system's chirality.  
As the system size increases, the critical point $h_c$ gradually converges toward a range between 0.95 and 1.

The local Chern marker  is directly related to the physically measurable orbital magnetization, $\mathcal{M}_\text{orbital}(r)$. For a generic system, orbital magnetization has two parts--nontopological  and topological magnetization. In our case, due to the particle-hole symmetry of the problem, the nontopological contribution vanishes. 
 To confirm this we also compute non-topological orbital magnetization numerically (see Appendix~\ref{appendix:magnetization}) and find it to be identically zero.
This leaves the total orbital magnetization directly proportional to the local Chern marker:
\begin{align}
    \mathcal{M}_\text{orbital}=\frac{e}{h c}\mu M. \label{eq:Morb}
\end{align}
Here $e$ is the electron charge and $\mu$ is the chemical potential measured from $\mu=0$ at half filling. (In this formula with only the topological magnetization, the arbitrariness of a reference energy $\mu_0$ can 
rearrange the contributions of topological and nontopological terms to the orbital magnetization, but the full orbital magnetization is a physically measurable quantity, invariant to such changes, and will still be given by Eqn.~\ref{eq:Morb}.) Due to the prefactor involving $\mu$ this magnetization vanishes when $\mu$ is precisely zero. However, tuning $\mu$ by any small amount will manifest the nonzero orbital magnetization. For example, even for a uniformly gapped Chern insulator in the Haldane model limit, tuning $\mu$ within  the bulk gap (by filling chiral edge states) will produce the nonzero orbital magnetization. We thus expect that in experimentally relevant settings, the orbital magnetization will be nonzero.

To further investigate the topological chirality, we now turn to compute the probability currents of low-energy occupied states to see whether a similar sign flip occurs as in the Chern marker computed from all occupied states.

\subsection{Circulating low energy currents}

Since the chiral edge modes in the Haldane model reflect the system’s topology, we now aim to examine the probability current circulation in low energy states to gain insight into the topological characteristics in the presence of a single defect. 
Consistent with the results from the local Chern marker, we find that the chirality of the system, characterized by the direction of the probability current in low-energy states, undergoes a reversal as $h$ increases. Interestingly, this indicator exhibits a slightly different $h_c$, and its convergence is much faster and more stable. %

For simplicity we define the current as the  probability current, but of course including the electron charge gives the experimentally measurable electronic current.
For any   state $\psi$ the current can be computed for each directed edge $a\rightarrow b$: 
\begin{align}
    J_{ab} &= i \left( \psi_a^* H_{ab} \psi_b - \psi_b^* H_{ab} \psi_a \right) 
    = 2\operatorname{Im}[\psi_b^* H_{ab} \psi_a]
    \label{eq:Jba}
\end{align}
where $H_{ab}$ represents the hopping term from point $a$ to $b$.

We compute the current in the occupied low-energy states (states with energy just below the chemical potential $\mu=0$).
Unlike the case of a uniformly bulk-gapped Chern insulator, there is no universal definition of "low energy" in the present system. However we find that the choice of definition does not modify the presence of the chirality reversal transition, nor does it systematically impact  the value of $h_c$. 
To smoothly isolate low-energy, occupied states, we use a Gaussian filter function: $f(\epsilon) = (2 / (\sqrt{2\pi}\sigma))\,e^{-(\epsilon - E_0)^2 / (2\sigma^2)}$
restricted to $\epsilon\leq0$, with $E_0=0$ and $\sigma=\alpha/L$. 
In a finite 2D system, the Dirac cones exhibit finite-size gaps that scale as $1/L$, which helps to distinguish effectively low-energy states. To reflect this, we define the Gaussian width as $\alpha/L$ and thus the parameter $\alpha$ sets the low-energy scale, which allows us to optimize the definition of "low energy" by minimizing finite-size effects. The inset of Fig.~\ref{fig3} (discussed below) shows results for $\alpha= 0.5, 1, 2, 4$, illustrating the effect of varying the Gaussian width $\alpha/L$. The features remain robust despite some variations across different values of $\alpha$. Among these, $\alpha=2$ (Gaussian width $2/L$) yields minimal artifacts and appears to be a physically reasonable choice throughout our calculations.

To obtain  the distribution of probability currents in low-energy states, we apply the Gaussian filter function and then calculate the probability current defined by Eq.~\ref{eq:Jba} to directly plot the current on each bond in the central region for clear visualization. The subfigures in Fig.~\ref {fig1} show the results for $h=0.7$ and $h=1.3$, while the complete range from $0.7$ to $1.5$ is provided in the Appendix~\ref {appendix:visualization}. The key observation is that as $h$ increases, the currents outside the defect core reverse their circulation from counterclockwise to clockwise.%

We define circulation flow $I_\text{circ}$ as the sum of low energy currents on all edges intersecting a radial line from the origin. Counterclockwise circulation is  taken  positive. By current conservation, the circulation flow is independent of the choice of the radial line; for concreteness, we choose the positive x-axis (grey dashed line in Fig.~\ref{fig1}):
\begin{equation}
    I_{\text{circ}} \equiv \sum\limits_{a\rightarrow b} J_{ab}, \quad
    \text{edge } a\rightarrow b \text{ crosses line } y=0,x\geq0
    \label{eq:I_circ}
\end{equation}

Fig.~\ref {fig1}(b) shows $I_\text{circ}$ in low-energy states across the positive x-axis for $\alpha = 2.0$ %
over different system sizes. All cases exhibit a zero-crossing point around $0.9$, as the current flow is counterclockwise before it and becomes clockwise afterward. Below $h_c$, the chirality of the system is consistent with that of the uniform Haldane model: When the imaginary NNN hopping is defined as $+i$ in the counterclockwise direction, the Chern number is $+1$, corresponding to counterclockwise edge-state currents. Above $h_c$, the circulation reverses.

While both probes indicate a clear chirality reversal, they exhibit slightly different behavior. The top right panel in Fig.~\ref {fig3} compares the transition point $h_c$ obtained from local Chern marker $M$ and from low energy current flow $I_\text{circ}$ with different parameters (size of edge vs bulk region and low energy filter Gaussian width $\sigma=a/L$, respectively). The marker probe places the transition at $h_c\approx1$ and is more sensitive to finite-size effects (as illustrated in Fig.~\ref {fig3} main panel). In contrast, the current-flow probe converges to $h_c\approx0.9$ with stable behavior even for smaller system sizes (as shown in Fig.~\ref{fig1}b).

\section{Local signatures of chirality reversal:  defect-core currents, toy model, and impurity projected T-matrix}
\label{sec:locality}

\begin{figure*}[t]
  \centering
    \begin{overpic}[width=0.49\textwidth]{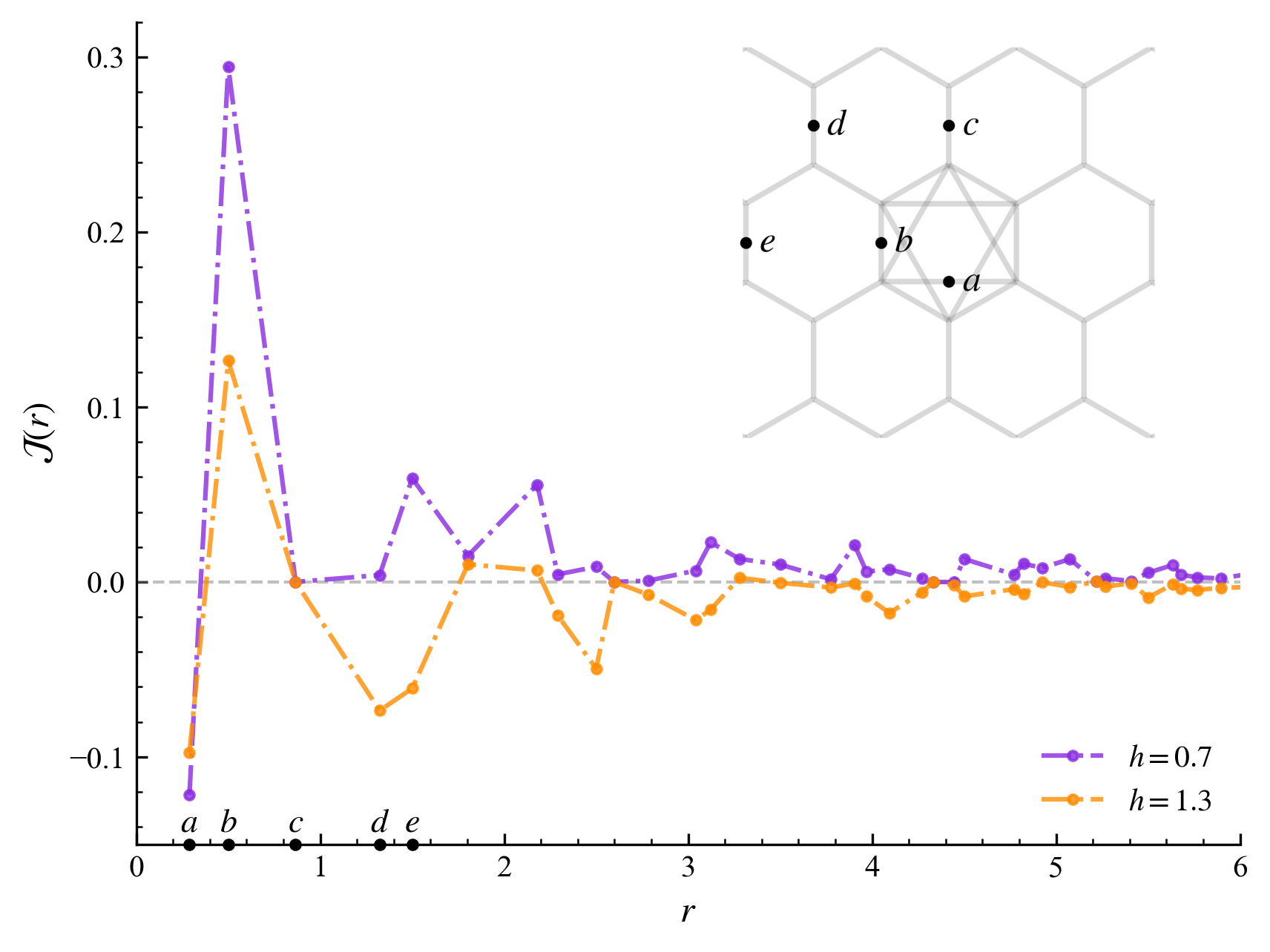}
    \put(3,75){\small\text{(a)}}
  \end{overpic}
  \hfill
  \begin{overpic}[width=0.49\textwidth]{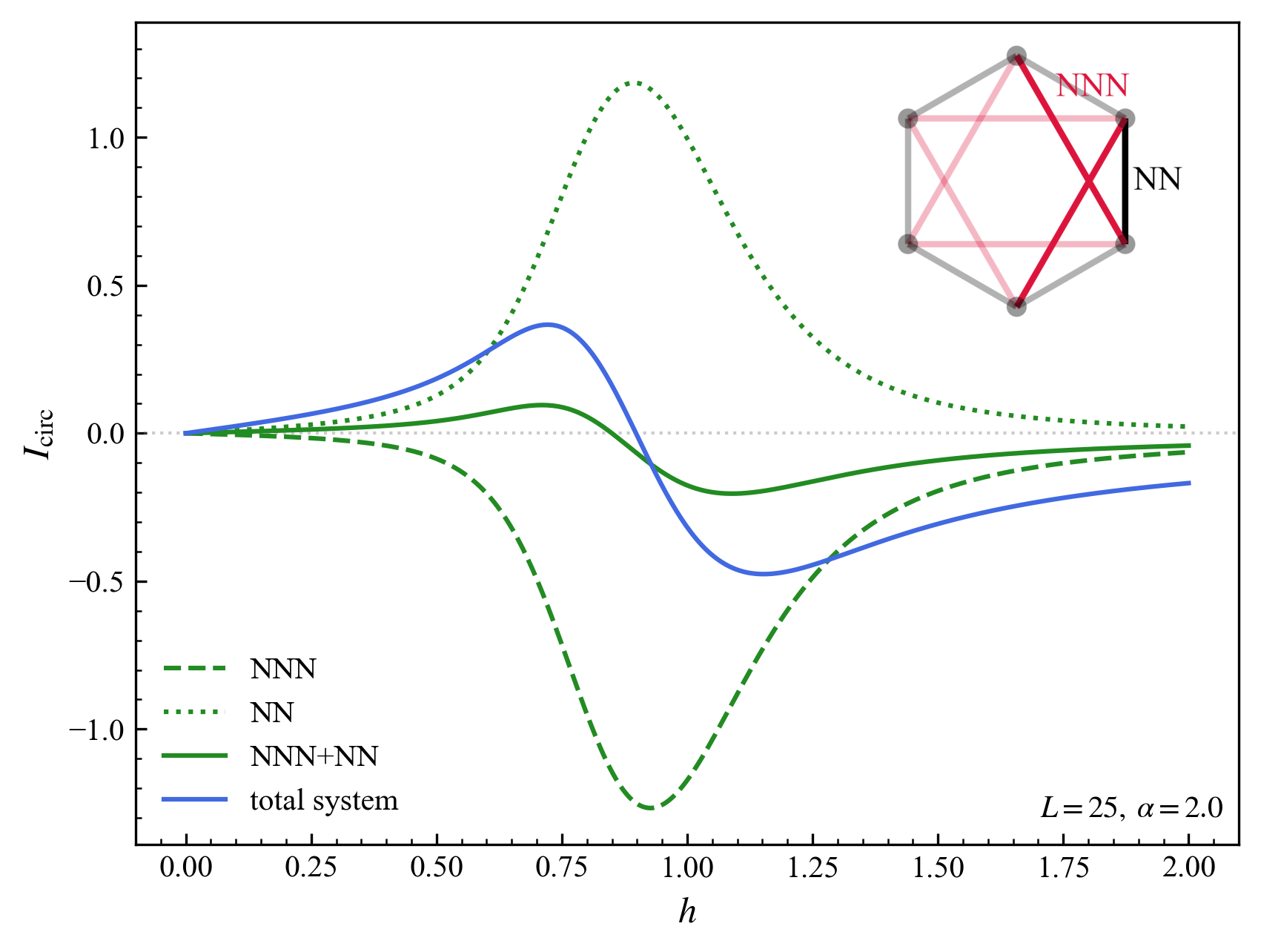}
    \put(3,75){\small\text{(b)}}
  \end{overpic}
  \caption{Localized and extended currents and their chirality reversals. (a) Distribution of $\hat{\theta}$ component of bond current, $\mathcal{J}(r)$ (Eq.\ref{eq:Jr} ),  as a function of radial distance $r$ for two values of $h$, below and above $h_c$.   
  The smallest $r$ points, $a,b,c,d,e$, are shown for clarity.
  Current at $a$ ($b$) is always negative (positive) respectively (see right panel). Bond $c$ is radial hence has $\mathcal{J}=0$.
  The $\hat{\theta}$ current at $d$,$e$ flips sign between $h=0.7$ and $h=1.3$. 
  Results are calculated for $L=25$, $\alpha=2$. 
  The low energy current is primarily concentrated near the central defect area and decays with distance. 
  The sign flip indicating chirality reversal is visually apparent for $r>1$.
  (b) Locality of sign flip of low energy circulating current $I_\text{circ}$. $I_\text{circ}$ across the full system (blue line) and $I_\text{circ}$ on bonds of the central hexagon defect contributing to $I_\text{circ}$  (highlighted in inset): 
  two next-nearest-neighbor (NNN) imaginary-hopping bonds (dashed green line); nearest-neighbor (NN) real-hopping bond (dotted green line); and their sum (solid green line).
The near cancellation of NNN negative currents and NN positive currents enables the net NN+NNN current on defect core bonds to show a chirality reversal at $h\approx 0.85$, near the total current reversal at $h\approx 0.9$. 
  }
\label{fig4}
\end{figure*}

In this section, we analyze the chirality reversal across different spatial length scales. We begin by showing the current distribution of low-energy states over different regions of the sample, and then focus on the features at the defect core. We interpret the numerical results in terms of a toy model on a single defect. We then relate the toy model to the impurity projected T-matrix.

\subsection{Spatial distribution of currents showing chirality flip}

To visualize the radial distribution of the circulating current, we adopt a slightly modified definition of current circulation that balances the absence of circular symmetry in the sample with the need to focus on "circulation."
The definition we use for the radial distribution $\mathcal{J}(r)$ is
\begin{equation}
    \mathcal{J}(r) \equiv \frac{1}{N(r)} \sum_{r} J_{ab}\hat{n}_{a\rightarrow b}\cdot\hat{\theta}_{ab}
    \label{eq:Jr}
\end{equation}
Mathematically, it represents the average of currents in tangential directions at a distance $r$, where $r$ is the distance from the midpoint of edge $a\rightarrow b$ to origin, as shown in the subfigure in Fig.~\ref {fig4}(a). $N(r)$ is the number of directed edges at a same distance $r$, $\hat{n}_{a\rightarrow b}$ is the unit vector of the bond $a\rightarrow b$, and $\hat{\theta}_{ab}$ is the polar angle unit vector in polar coordinates at the midpoint of edge $a\rightarrow b$. Note that the low energy states are consistently selected using the same Gaussian filter before computing $\mathcal{J}(r)$.

Fig.~\ref{fig4}(a) shows $\mathcal{J}(r)$ at $h=0.7$ and $h=1.3$, below and above $h_c$. The current intensity  decreases with distance, indicating that the low energy currents are relatively concentrated near the center; the associated magnetic moments, given by the current times the enclosed area, also decay with $r$ albeit slowly. Except at the central region, the current direction visually reverses between 0.7 and 1.3, suggesting a chirality reversal. This behavior is consistent with the analysis in the previous section, where the current flow $I_\text{circ}$ is defined by Eq.~\ref{eq:I_circ}.

To compare the contribution to the circulating current from the defect core versus from the rest of the system, we now return to $I_\text{circ}$. 
As shown in panel \ref{fig4}(b), the low-energy probability flow along the defect-core imaginary NNN hoppings (the added Haldane-type defect) is always clockwise, with its strength first increasing and then decreasing. In contrast, the flow along the defect-core real NN hopping (the edges of the central hexagon) is consistently counterclockwise, and its strength also follows a similar trend. The two contributions (the entire defect core region) nearly cancel each other out, with a sign change occurring near $h_c$. Since the NN+NNN bonds of the defect hexagon do not form a complete system with conserved current, unlike the total current  the NN and NNN currents can depend on the reference axis. Here we show the results calculated for currents flowing across  the positive x-axis. However the plot for  currents flowing across y-axis is visually nearly identical. The sign change occurs at a critical $h$ only a bit smaller than the total system $h_c$.

This interesting observation indicates that the local behavior near an isolated defect  may  already predict the chirality change. When $h$ is either small or large, low-energy states might be sparsely distributed in the defect region, or they exist but their current contributions are negligible. 
Even when restricted to NN or NNN bonds within the defect core (dotted or dashed green lines), the current exhibits a broad peak around $h\approx0.6-1.3$, near the transition point $h_c$. This is in line with the accumulation of low energy states near the gap-closing transition and demonstrates that the effect is observable even at the defect core.

\subsection{Defect scale toy model for chirality reversal}

To explain the cancellation between the NNN and NN core-defect currents we consider a toy model consisting of a wavefunction $\psi$ that is a bound state with substantial amplitude on the six sites of the defect hexagon.
The assumption of substantial amplitude on the defect hexagon implies that $\psi$ should be a defect bound state.
We rewrite the probability current (Eq.~\ref{eq:Jba})   from site $a$ to $b$ for state $\psi$ explicitly as:
\begin{align}
    J_{ab} &= 2\,\operatorname{Im}\left[\psi_b^* H_{ab} \psi_a\right]\\
    &= 2\,\operatorname{Im}\left[|\psi_b| e^{-i\theta_b}\;|t_{ab}| e^{i\phi_{ab}} \; |\psi_a| e^{i\theta_a} \right]\\
    &=2\,|t_{ab}||\psi_b||\psi_a| \sin\left(\theta_a-\theta_b+\phi_{ab}\right)
    \label{eq:phase_current}
\end{align}
where the hopping term from $a$ to $b$ is $H_{ab} = |t_{ab}| e^{i\phi_{ab}}$ and $\theta_{a,b}$ denotes the phase of $\psi$ at site $a$ or $b$.

\begin{figure}[b]
\centering
\begin{minipage}{0.22\linewidth}
\centering
\includegraphics[width=\linewidth]{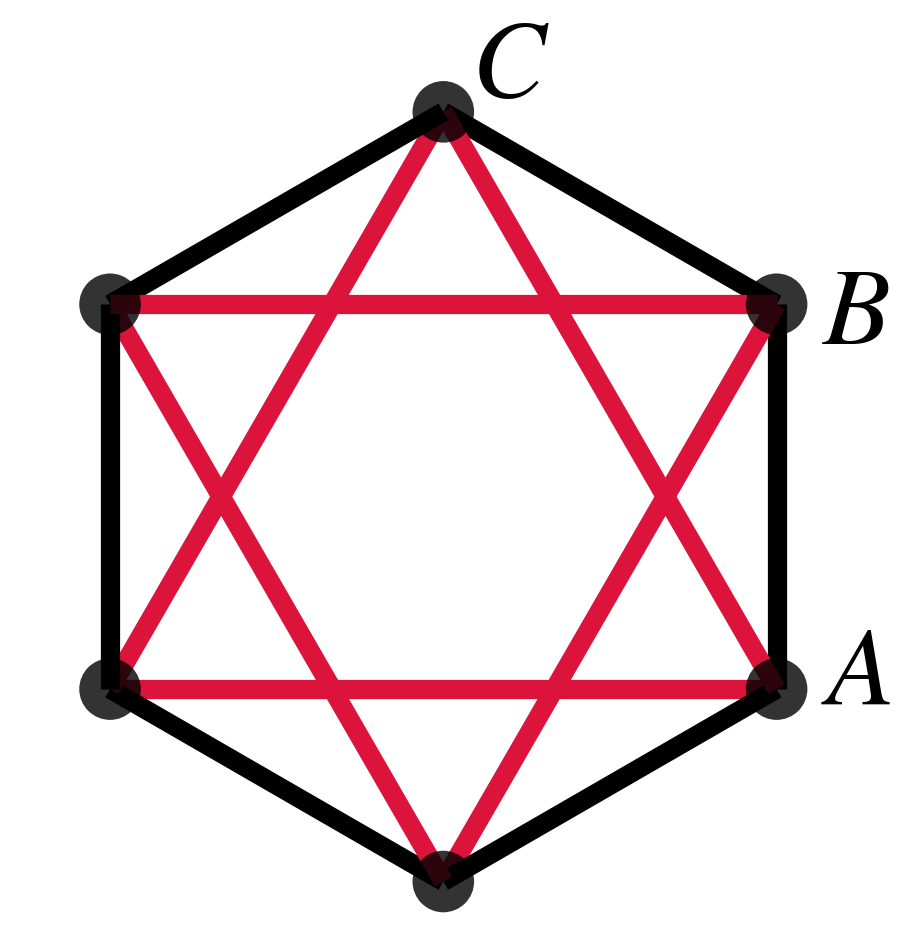}
\label{fig:table}
\end{minipage}
\hspace{2pt}
\begin{minipage}{0.65\linewidth}
\centering
\begin{tabular}{|c|c|c|c|}
\hline
$\theta\equiv\theta_A-\theta_B$ & $J_{AB}$ & $J_{AC}$ &
$I_\text{circ}^\hexagon$
\\ \hline
0, $\pi$ & 0 & $2h$ & $4h$
\\ \hline
$\pi/3$, $2\pi/3$ & $-\sqrt{3}$ & $-h$ & $-\sqrt{3}-2h$ 
\\ \hline
$4\pi/3$, $5\pi/3$ & $\sqrt{3}$ & $-h$ & $\sqrt{3}-2h$ 
\\ \hline
\end{tabular}
\label{tab:phase_current}
\end{minipage}
\caption{Single hexagon toy model of low energy wavefunction phase windings. Left: Counterclockwise labeling of sites A, B, C on the Haldane hexagon defect. Right: 
Current $J$ on NN bonds ($J_{AB}$) and NNN bond ($J_{AC}$), and resulting NN+NNN $I_\text{circ}^\hexagon = J_{AB}+ 2J_{AC}$, 
as functions of wavefunction phase difference $\theta$ between neighboring sites. Currents are expressed in units of $|\psi|^2$. Note that $J_{AC}$ involves the imaginary hopping $h$. The low energy states have $\theta=4\pi/3,5\pi/3$, giving a sign flip for $I_\text{circ}'$ at a critical $h_c^\hexagon=\sqrt{3}/2$. }
\label{fig5}
\end{figure}

Let A, B, C be three consecutive vertices of the hexagon in counterclockwise order (Fig.~\ref{fig5}). Then $H_{AB}=-1$ and $H_{AC}=ih$.
We define the phase difference between states at adjacent points as $\theta\equiv\theta_A-\theta_B=\theta_B-\theta_C=\frac{1}{2}(\theta_A-\theta_C)$. This clockwise sign convention for $\theta$ phase differences gives positive $\theta$ for states of the model with the most-negative energies (Eq.~\ref{eq:etheta}).
In the toy model we  assume  60-degree rotational symmetry around the hexagon center which labels states by their angular momentum proportional to $-\theta$.  Without rotation symmetry one could still define $\theta$ as an average phase difference between successive hexagon sites. 
With rotation symmetry the phase winding $\theta$ takes integer values in units of $2\pi/6=\pi/3$. The corresponding currents computed in each state of a given $\theta$ are listed in the Table of Fig.~\ref{fig5}. 

Our numerical results appear to closely match the bottom row of the table: the currents on defect NN bonds is always positive (counterclockwise), while those on defect NNN bonds is always negative (clockwise). Let us then consider low-energy bound states with $\theta$ roughly within the range $4\pi/3$ to $5\pi/3$, i.e.\ near $3\pi/2\equiv -\pi/2$. For such a low energy bound state we have $J_{\text{NNN}} = 2J_{AC}=-2h|\psi|^2$ and $J_{\text{NN}}=J_{AB}=\sqrt{3}|\psi|^2$.  Calculating the full circulating current flow within the 6-site toy model, these exactly cancel each other out at $h=h_c'$ with $h_c'=\sqrt{3}/2 \approx 0.87$.

Indeed our numerical results show that the low energy states are well described by this range of phase windings $\theta$ (see Appendix~\ref{appendix:phase}). Even though our sample does not have strict rotational symmetry, most low-energy states are concentrated around $\theta$ values of $4\pi/3$ and $5\pi/3$. 

To  understand why the low energy states exhibit this range of phase differences between adjacent vertices, it is enough to again analyze a simple toy model consisting of just the six sites forming the central hexagon (see Appendix ~\ref{appendix:phase}). Assuming that the NN hopping is $t=-1$ and NNN hopping is $ih$ in the counterclockwise direction as before, then the six eigenstate energies are 
\begin{align}
E_\hexagon(\theta)=-2\cos{\theta}-2h\sin{2\theta}
\label{eq:etheta}
\end{align}
 where $\theta$ is an integer multiple of $2\pi/6$. The low-energy states (those closest to zero energy) have energies $-\sqrt{3}h+1$ or $\sqrt{3}h-1$ for the relevant range of $0<h<\sqrt{3}$, with phase difference $\theta$ of $4\pi/3$ or $5\pi/3$, respectively. More generally for either sign of $h$, the relevant low energy states cluster around $\theta \sim -\text{sgn}(h) 3\pi/2 $. 
 This toy model thus exhibits a chirality reversal transition at $h'_c=\sqrt{3}/2 \approx 0.87$ arising from a sign flip of current flow in the low energy bound states. %
 This feature is again consistent with the full numerical computations showing chirality reversal of the core defect bond currents at $h\approx 0.85$ (Fig.~\ref{fig4}(b)).

\subsection{Critical point from impurity projected T-matrix}

To complement the toy model, a full IR limit description of the chirality reversal critical point  can be obtained by considering the impurity projected T-matrix of the system \cite{slager_impuritybound_2015, queiroz_ring_2024}.   This impurity projected T-matrix is closely related to the Dirac cone projected T-matrix discussed in Sec.~\ref{sec:T-matrix} and Ref.~\cite{neehus_genuine_2025}. Both approaches involve a low energy or $E=0$ projection. However, the Dirac cone projection relies on momentum space. This enables a description of the effective Dirac cone mass term generated by the impurity, but correspondingly does not directly access the real space wavefunctions involved in the resonance. In contrast, the  impurity projected T-matrix is constructed in real space. This provides access to the real space wavefunctions of the  impurity quasi-bound states  responsible for the resonance that creates the critical point.

At the  critical point  the model undergoes a gap closing transition which  can be obtained from the pole of the zero energy T-matrix. The pole arises from the bound state(s) $|\psi\rangle$ obeying
\begin{align}
    |\psi\rangle = G(0)V_{\hexagon'} |\psi\rangle
\end{align}
where $G(0)$ is the Green's function of $H_0$ (graphene) at zero energy.
The term bound state comes from the fact that a pole of the T-matrix corresponds to a pole of the full Green's function.
For an insulator, within the gap, this gives a bound state. In the present case the clean system Dirac cones have no gap so the resulting states are quasi-bound states which have amplitude near the defect and remain spatially extended. 
Generically $ G(0) V_{\hexagon'}$ is a non-local operator and it might appear difficult to solve the above eigenvalue equation. 
However, we will see that  $V_{\hexagon'}$ being a local operator, we can solve the eigenvalue equation by projecting it onto the defect eigenspace and then  the full wavefunction can be obtained from the projected solution.  

Let us start by defining the projector $P_{\hexagon'}$, 
which projects onto the non-zero eigenspace of $V_{\hexagon'}$. We obtain $P_{\hexagon'}$ by diagonalizing $V_{\hexagon'}$,
\begin{align}
    &P_{\hexagon'}=\sum_{S=A,B}\sum_{m=\pm} P_{S m}~~~ ,~~~  P_{S m}= |m\rangle_{S~S}\langle m|
\end{align}
where $|\pm\rangle_S$
are the positive and negative chirality eigenstates of $V_{\hexagon'}$ with negative and positive eigenvalues, respectively: 
$V_{\hexagon'}|\pm\rangle_S = \mp \sqrt{3} h |\pm\rangle_S$. 
The  wavefunction for $|\pm\rangle_S$ has the same form for both of the sublattices $S=A,B$, and is
\begin{align}
    |\pm\rangle_S=\frac{1}{\sqrt{3}}\left(1, \phi_0^{\pm 1},\phi_0^{\mp 1}\right), \quad \phi_0=e^{2 \pi i/3}
\end{align}
on the three sublattice-$S$ sites of the hexagon $\hexagon'$ listed in counterclockwise order. Here 
$(\phi_0)^{\pm 1}=\exp(\pm 2 \pi i/3)$ give the complex phase and its complex conjugate. 
Note $P_{\hexagon'}^2=P_{\hexagon'}$ and $P_{\hexagon'}V_{\hexagon'} =V_{\hexagon'} P_{\hexagon'} =V_{\hexagon'}  $.
Denoting Pauli matrices $\mu$ and $\sigma$ acting on the chirality and sublattice indices, the projected $V_{\hexagon'}$ can be decomposed as 
\begin{align}
    P_{Sm}V_{\hexagon'}P_{S'm'}=-\sqrt{3}h\sigma^0_{ss'}\mu^z_{mm'}.
    \label{eq:PV}
\end{align}
with $\sigma^0$ being the identity matrix. 

We  first solve for the projected part of the wavefunction $|\psi\rangle$,
\begin{align}
    P_{\hexagon'}|\psi\rangle=P_{\hexagon'}G(0)V_{\hexagon'}|\psi\rangle\equiv P_{\hexagon'}G(0)V_{\hexagon'}P_{\hexagon'}|\psi\rangle.
    \label{eq:projected pole}
\end{align}
Once $P_{\hexagon'}|\psi\rangle$ is known, the rest of the wavefunction $\left(1-P_{\hexagon'}\right)|\psi\rangle$ can then immediately be obtained by applying $G(0)V_{\hexagon'}$, as follows,
\begin{align}
    \left(1-P_{\hexagon'}\right)|\psi\rangle=(1-P_{\hexagon'})G(0)V_{\hexagon'}P_{\hexagon'}|\psi\rangle. 
    \label{eq:QV}
\end{align}
Hence to determine the full bound state wavefunction, we need only solve for the projected part.

The projected Eq.~\ref{eq:projected pole} can be solved by finding eigenvalues of the rank-4 operator 
$P_{\hexagon'}G(0)V_{\hexagon'}P_{\hexagon'}=\left(P_{\hexagon'}G(0)P_{\hexagon'}\right)\left(P_{\hexagon'}V_{\hexagon'}P_{\hexagon'}\right)$. 
The lattice Green's function $G(0)$ is non-local  and generically has non-zero element between any two $A,B$ sites. However, for computing $P_{\hexagon'} G(0) P_{\hexagon'}$ it is enough to consider only the six site Green's function on the defect hexagon which has only two distinct matrix elements, namely nearest neighbor and third nearest neighbor. In the basis of the six sites, $G$ (with $E=0$ implied here and below) then becomes a real symmetric matrix with nonzero matrix elements
\begin{align}
G_{i,i\pm1}=g_1=\frac{1}{3}, \quad G_{i,i+3}=g_3=-\frac{\sqrt{3}}{2\pi}
\end{align}
with $i$ interpreted cyclically on the 6 sites (see for example Eq. S5 of Ref. \cite{kot_band_2020}).
The nonzero entries are $g_1$ and $g_3$, which are the nearest-neighbor and third-neighbor matrix elements of the honeycomb lattice Green's function.
Note that the second neighbor matrix element involves the same sublattice and hence vanishes, $g_2=0$, since the honeycomb lattice is bipartite.

Projecting $G$ into the defect subspace gives
\begin{align}
    P_{ Sm} G P_{S'm'}
    &= g_{31} \left(      \phi_{x}\sigma^x_{SS'}\mu^0_{mm'}-\phi_{y}\sigma^y_{SS'}\mu^z_{mm'}\right)
    \\
    g_{31}&=g_3-g_1= -\left( \frac{\sqrt{3}}{2\pi} + \frac{1}{3} \right)
    \end{align}
where $\mu^0$ is the identity matrix and the numbers $\phi_{x}=-1/2$ and $\phi_{y}=\sqrt{3}/2$ are the real and imaginary parts of the complex phase 
$\phi_0=\exp(2 \pi i/3)$. %
Note that $\phi_0$ determines the chirality of the impurity projected eigenstates.
Multiplying by Eq.~\ref{eq:PV} gives the defining equation
    \begin{align}
    P_{Sm}G V_{\hexagon'}  P_{S'm'}
    &=\sqrt{3} g_{31} \ h \  \mathcal{O}_{SS',mm'}
    \\
    \mathcal{O}_{SS',mm'}&=
      -\phi_{x}\sigma^x_{SS'}\mu^z_{mm'}+\phi_{y}\sigma^y_{SS'}\mu^0_{mm'}
\end{align}

We are now ready to solve this equation to identify the pole responsible for the critical point.
The matrix $\mathcal{O}$ has eigenvalues $\pm 1$, each occurring with a twofold degeneracy associated with the $\mu^z$ positive and negative chirality states.
Among these eigenvalues the pole of the zero energy T-matrix is determined by those with opposite sign from $h$. %
These two ``$-\text{sgn}(h)$''  eigenvalues produce a pole at $h_c$ which is simply determined by $\sqrt{3} |g_{31}| h = 1$ giving
\begin{align}
    |h_c|= \frac{1}{\sqrt{3}|g_{31}|}\approx0.948.
\end{align}
This is the same pole as found in the Dirac cone projected T-matrix, but here determined through the full T-matrix with a projection to the impurity states enabling  interpretation of the responsible wavefunctions.
Computing the  eigenstates of $\mathcal{O}$ corresponding to the $\text{sgn}(-h)$ eigenvalues, we find that the phase differences between successive $A,B$ sites are $\theta=\theta_A-\theta_B=5\pi/3, 4\pi/3$ 
for the two degenerate states that produce the pole. The  state with $5\pi/3$ phase difference is a positive chirality state since it is a superposition of $|+\rangle_{A}$ and $|+\rangle_B$. Similarly, the state with $4\pi/3$ phase difference is a negative chirality state.  %
Equivalently, the corresponding eigenvalues under the threefold rotation symmetry are $-2\theta = \pm 2\pi/3$ mod $2\pi$ for $\theta=5\pi/3, 4\pi/3$ respectively.

\subsection{Comparison of toy model sign flip $h_c'$ and global chirality reversal $h_c$}

We note two  features that relate  the toy model's defect-scale sign change at $h_c'$ to the global chirality reversal phase transition produced by the T-matrix resonance  at $h_c$.

First, we observe that $h_c$ and $h_c'$ are produced by exactly the same two defect-bound states. 
By analyzing the impurity projected T-matrix we showed that the  $\theta=5\pi/3, 4\pi/3$  states are responsible for the resonance that produces the global chirality reversal phase transition. 
These are  the same states observed numerically and determined analytically by the toy model (last row of the Fig. \ref{fig5} Table). In the T-matrix they appear as full bound states, not just restricted to the defect core, by application of $G(0)V_{\hexagon'}$ as in Eq.~\ref{eq:QV}. 
Similarly, though in the toy model the energy of these states is not  zero at the transition,  the full T-matrix computation  indicates that they are indeed zero energy bound states at the critical point.
Regardless, both in the defect-scale toy model and in the $E=0$ T-matrix, these two particular bound states states are responsible for the chirality reversal.

Second, we note an intriguing relation between the values of the $h_c$ critical point exhibited by the impurity projected T-matrix and the $h_c'$  sign change of the defect scale toy model. 
Consider $h_c$. The first and third neighbor matrix elements of the Green's function, $g_1$ and $g_3$, are the only nonzero matrix elements that are involved within the 6-site impurity projected T-matrix, since the impurity potential here only involves 6 sites. Now observe that if the third neighbor term in the Green's function $G_{i,i\pm3} = -\sqrt{3}/2\pi \approx  -0.28 $ instead took the same value as the (negated) NN term $-G_{i,i\pm1} = -1/3\approx- 0.33$, then the transition points would both occur at the same $h_c'=\sqrt{3}/2$.  The difference between $g_3$ and $-g_1$ produces the shift from $h_c'$ to $h_c$. If instead of the infinite system we had considered a finite system consisting of a single hexagon, its Green's function would indeed have $g_1=-g_3$. %
In the present case these two matrix elements  are evidently sufficient to encode the relevant global information from the Green's function  within the defect-restricted zero-energy bound states. 
The shift of the critical point $h_c$ away from the defect-core chirality flip $h_c'$ then arises entirely due to the low energy long-range effects responsible for the nontrivial value of Green's function 3rd neighbor matrix element. Finally, this dependence itself relies on the particular microscopic structure of the impurity potential considered here.

\section{Discussion}

The chirality reversal transition is a topological phase transition between Chern number $C=1$ and $C=-1$ phases of the gapless Dirac cone system modified by a small density of Haldane hexagon magnetic impurities. Its appearance as a critical point at vanishing impurity density is consistent with the appearance of a resonance at the T-matrix pole. We find that this T-matrix pole zero energy bound state wavefunctions, as well as  certain features of the numerically computed chirality probes, 
show surprisingly local signatures.
When computed across the entire system, the circulating currents and the total Chern marker related to orbital magnetization both show chirality reversal transitions at a range of $h_c$ which is approximately similar to the $h_c$ describing the bulk Chern number flip and corresponding T-matrix pole.
However even when restricting to the Haldane hexagon defect core bonds, the circulating current still reverses its chirality, at a slightly smaller value $h_c'$, which can be fully captured in a simple 6-site toy model of the defect core.

This local signature is especially surprising given the fact that the chirality reversal transition relies on a \textit{dilute} small density of defects placed on top of the gapless Dirac cone background. At high defect densities the critical point disappears and chirality never reverses, as is well known for the uniform Haldane model where Chern number is simply set by the sign of $h$ with no chirality reversal at any nonzero $h$. In that sense the ``empty space'' between impurities might appear to play a key role.

Nevertheless we find that the currents within the defect hexagon reverse at a critical $h'_c$ that is only slightly smaller than the true $h_c$ phase transition of the full system. The relation between local and global probes of the chirality reversal is further supported by the  bound states responsible for the transition. The 6-site  toy model 
 chirality reversal  at $h'_c=\sqrt{3}/2\approx 0.87$ 
 and the full impurity projected T-matrix pole at $h_c=(\sqrt{3}|g_{31}|)^{-1}\approx 0.95$ both arise via the pair of zero energy bound states with phase windings $\theta=5\pi/3,4\pi/3$, which can be interpreted as the pair of opposite chirality states with ``positive'' angular momentum. These states 
locally produce the defect core chirality flip while also globally
producing the T-matrix resonance. The resulting T-matrix resonance and its associated divergence (an effective Dirac cone mass term that changes sign by diverging without crossing zero) enables the impurity-driven critical point to occur at  zero impurity density. 
The analysis suggests that the microscopic defect structure $V_{\hexagon'}$ captured by the defect-core toy model plays a role in creating the T-matrix resonance that generates the dilute impurity critical point.

Experimental implications of these results are clear: an inhomogenous TRB mass term arising in a Dirac cone system can change its effective sign as a function of its bare TRB strength. For example, adding spin-up magnetic impurities would produce positive orbital  magnetization for weak impurity scattering strength, as usually expected, but increasing the impurity scattering strength could suddenly reverse the orbital magnetization and the associated chiral currents --- producing negative electronic orbital magnetization even while the impurity spins remain ``up''.

Experimental settings where such magnetic impurity physics can arise include electronic Dirac cone materials as well as analogous Dirac cone systems of quasiparticles such as in superconductors or quantum spin liquids.
For the Kitaev honeycomb spin liquid  \cite{kitaev_anyons_2006}  Kondo spin impurities have already studied \cite{dhochak_magnetic_2010, keskiner_magnetic_2025} and the extension to frozen moments would produce  $V_{\hexagon'}$.
Regarding probes of chirality, it would be interesting to complement the usual bulk probes with  local magnetometry probes \cite{du_control_2017, tetienne_quantum_2017,mclaughlin_local_2023}. These could observe the enhanced magnetization near a single defect or cluster of defects, as is feasible in Moire systems.
 Local probes may be especially helpful since the chirality reversal does appear to require a low defect density (though we note that few-percent level impurity densities may be more common than expected \cite{kim_sharp_2025}). An experimental observation of such a reversal of the electronic orbital magnetization, occurring as a function of model parameters such as impurity strength, 
 would be a useful step towards understanding this unusual dilute impurity critical point.

\section{Acknowledgments}
We thank Xueda Wen, Mahmoud Asmar, and Kartiek Agarwal for useful discussions.
This work was supported by the U.S. Department of
Energy, Office of Science, Basic Energy Sciences, under
Early Career Award Number DE-SC0025478.

The data that support the findings of this article are openly available at the Georgia Tech Library Repository \cite{xu_data_2025}.

\appendix

\section*{APPENDIX}

\section{Current distributions}
\label{appendix:visualization}
\begin{figure*}[p]
    \centering
    \includegraphics[width=1\linewidth]{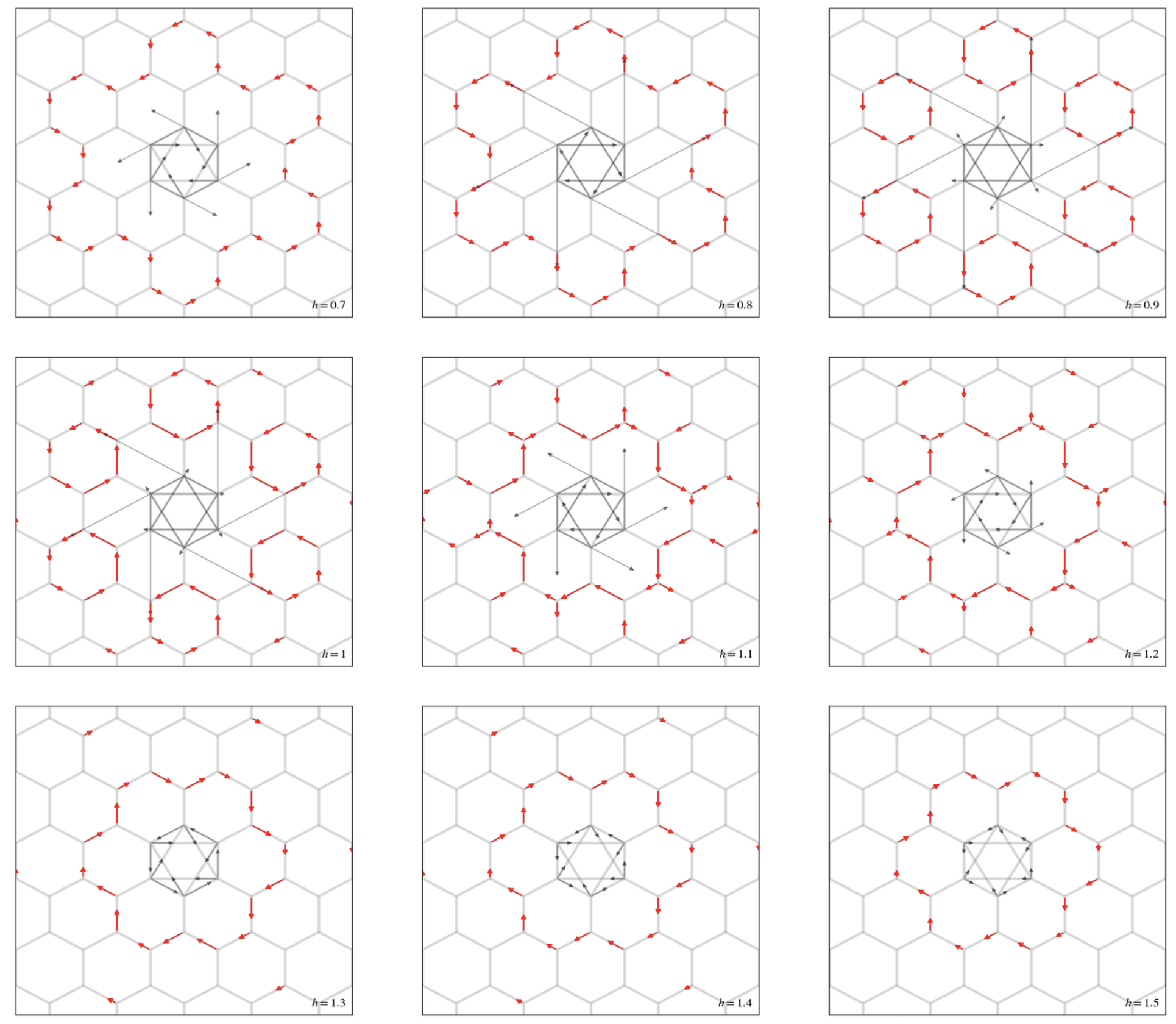}
     \caption{Currents of low-energy states from $h=0.7$ to $1.5$. The system size is $L=25$ and Gaussian function width is $\sigma=2/25=0.08$. Arrow lengths indicate current magnitudes, with tiny currents omitted. Currents on the defect hexagon bonds are shown in black, while all other currents are in red for better visibility. }
    \label{fig:visualization}
\end{figure*}

Fig.~\ref{fig:visualization} provides a complete visualization of how low energy current in the central region evolves from $h=0.7$ to 1.5. A clear reversal of current circulation is observed in the region away from the defect. Near the transition, a complex spatial distribution of current is observed.

\section{Local Chern marker}
\label{appendix:marker}
\begin{figure*}[t]
    \centering
    \includegraphics[width=1\linewidth]{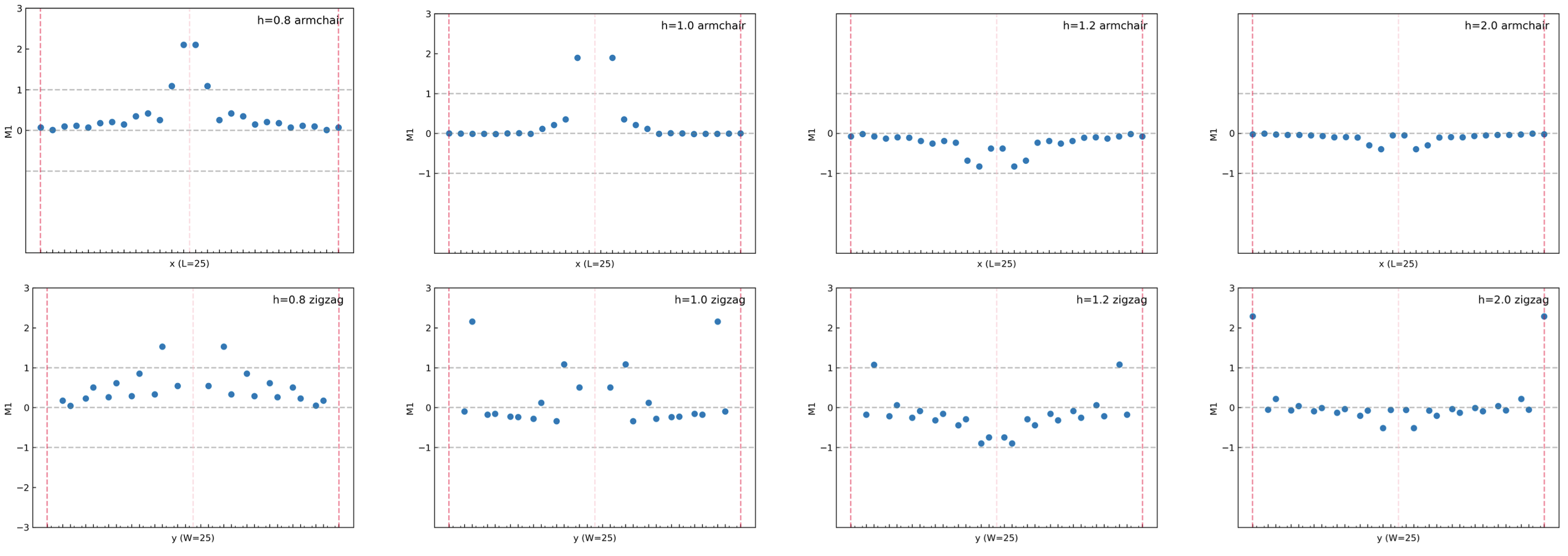}
    \caption{$M(r)$ values along 1D cuts through the central defect for $L=25$. Top row: horizontal path cutting through the sites just above the defect center ($y=\frac{1}{2}$). Bottom row: vertical path ($x=0$). The paths intersect the armchair and zigzag boundaries (red dashed lines), respectively. NNN hopping $h$  takes values 0.8, 1.0, 1.2, 2.0 for the four columns. The major x-axis ticks indicate lattice sites along the path, while the minor ticks denote other sites along this direction. The y-axis is fixed between -3 and 3, with values beyond this range omitted; sites near the boundary always contribute large opposite $M(r)$ such that the sum of $M(r)$ across the entire system identically vanishes.}
        \label{fig:marker_1d}
\end{figure*}

Figure ~\ref{fig:marker_1d} shows the real space distribution of the local Marker $M(r)$, as defined in Eq.~\ref{eq:chernmarker}, along a horizontal and a vertical path crossing through the central defect. Several spatial features can be observed: (1) the intensity is highest near the central defect area and decays with distance, with different behaviors along the horizontal and vertical directions; (2) near $h=1.0$, $M(r)$ at some points begins to change sign from positive to negative, indicating a reversal in chirality.

\begin{figure} 
  \centering
  \begin{adjustbox}{valign=t}
    \begin{overpic}[width=0.47\textwidth]{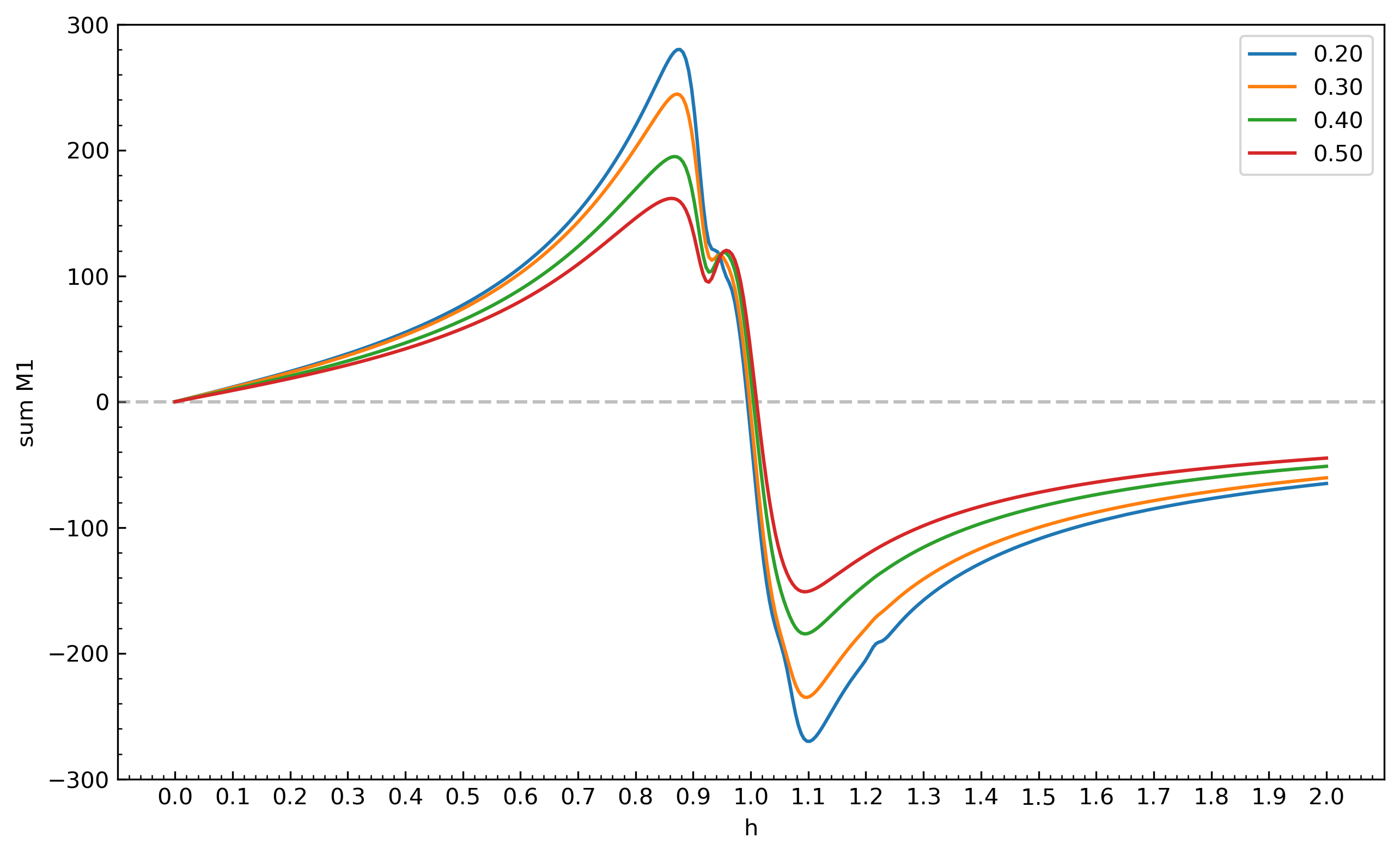}
    \end{overpic}
  \end{adjustbox}%
  \hfill
  \begin{adjustbox}{valign=t}
    \begin{overpic}[width=0.47\textwidth]{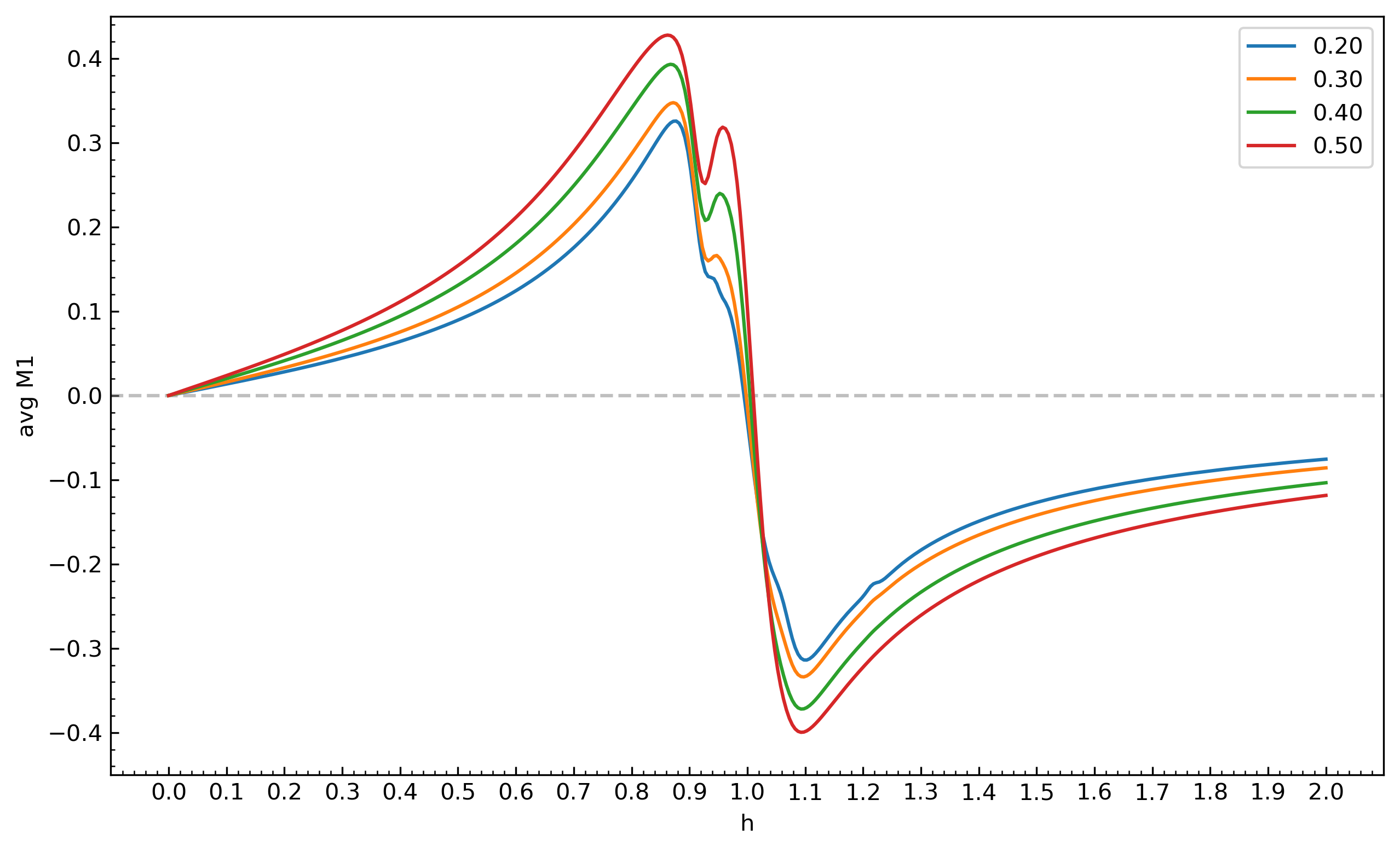}
    \end{overpic}
  \end{adjustbox}
  \caption{Summation and Average of $M(r)$ over different bulk regions at $L=25$. Different colors represent the linear fraction of sites belonging to edge region as in the main text.}
  \label{fig:sum_avg}
\end{figure}

Figure ~\ref{fig:sum_avg} shows a comparison between the summation  and average (total value divided by the number of points in the bulk region) of $M(r)$. It illustrates why the summation rather than the average is chosen as the  probe of interest to define $M$ in the main text, in this limit of a single defect. It better eliminates finite size effects and more smoothly reveals the topological features compared to the average value. It is also the more physical choice if we expect the magnetization to arise primarily from the defect as we do here, as opposed to through a uniform background magnetization density, since it then offers less dependence on the choice of "edge" region avoided by the bulk integration. 

\section{Non-topological magnetization}
\label{appendix:magnetization}
We numerically computed the total non-topological magnetization based on the definition introduced in Ref.~\cite{bianco_orbital_2013} and confirm it to be exactly zero, consistent with particle hole symmetry. 
For completeness we also computed further decompositions based on Ref.~\cite{bianco_orbital_2013} ,
$
    \mathcal{M}_\text{old}=\frac{1}{A_c}\operatorname{Im}\bra{r}PXHYP\ket{r}
$,  
and  
$
\mathcal{M}_\text{new}=\mathcal{M}_\text{LC}+\mathcal{M}_\text{IC}$, with $\mathcal{M}_\text{LC}=\frac{1}{A_c}\operatorname{Im}\bra{r}PXQHQYP\ket{r}$, $\mathcal{M}_\text{IC}=-\frac{1}{A_c}\operatorname{Im}\bra{r}QXPHPYQ\ket{r}
$.
We find that  $\mathcal{M}_\text{LC}$ and $\mathcal{M}_\text{IC}$ (which peak near the defect) are equal and opposite, so $\mathcal{M}_\text{new}$ exactly vanishes on every site. Correspondingly, the total sum of $\mathcal{M}_\text{old}$ also vanishes.

\section{Applicability of defect core toy model}
\label{appendix:phase}
\begin{figure} 
    \centering
    \includegraphics[width=0.49\textwidth]{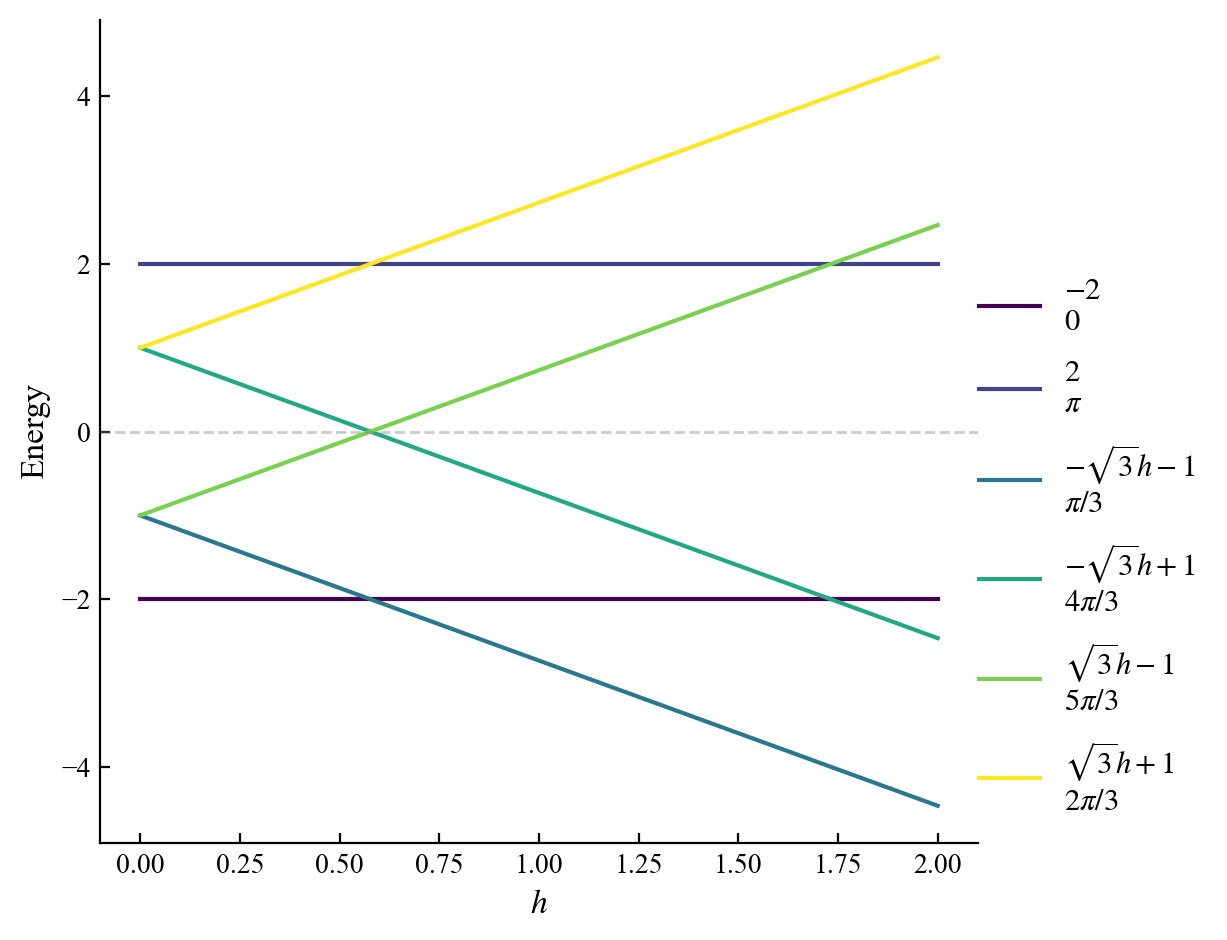}
    \caption{The energy as a function of $h$ in the  toy model consisting of 6 sites. The legend shows the corresponding eigenstate energy and the phase difference between adjacent sites. When $h<\sqrt{3}$, the   low-energy states have energies of $-\sqrt{3}h+1$ or $\sqrt{3}h-1$, with phase difference $\theta$ of $4\pi/3$ or $5\pi/3$, respectively.}
    \label{fig:toy}
\end{figure}

\begin{figure*}[t]
    \centering
    \includegraphics[width=1\linewidth]{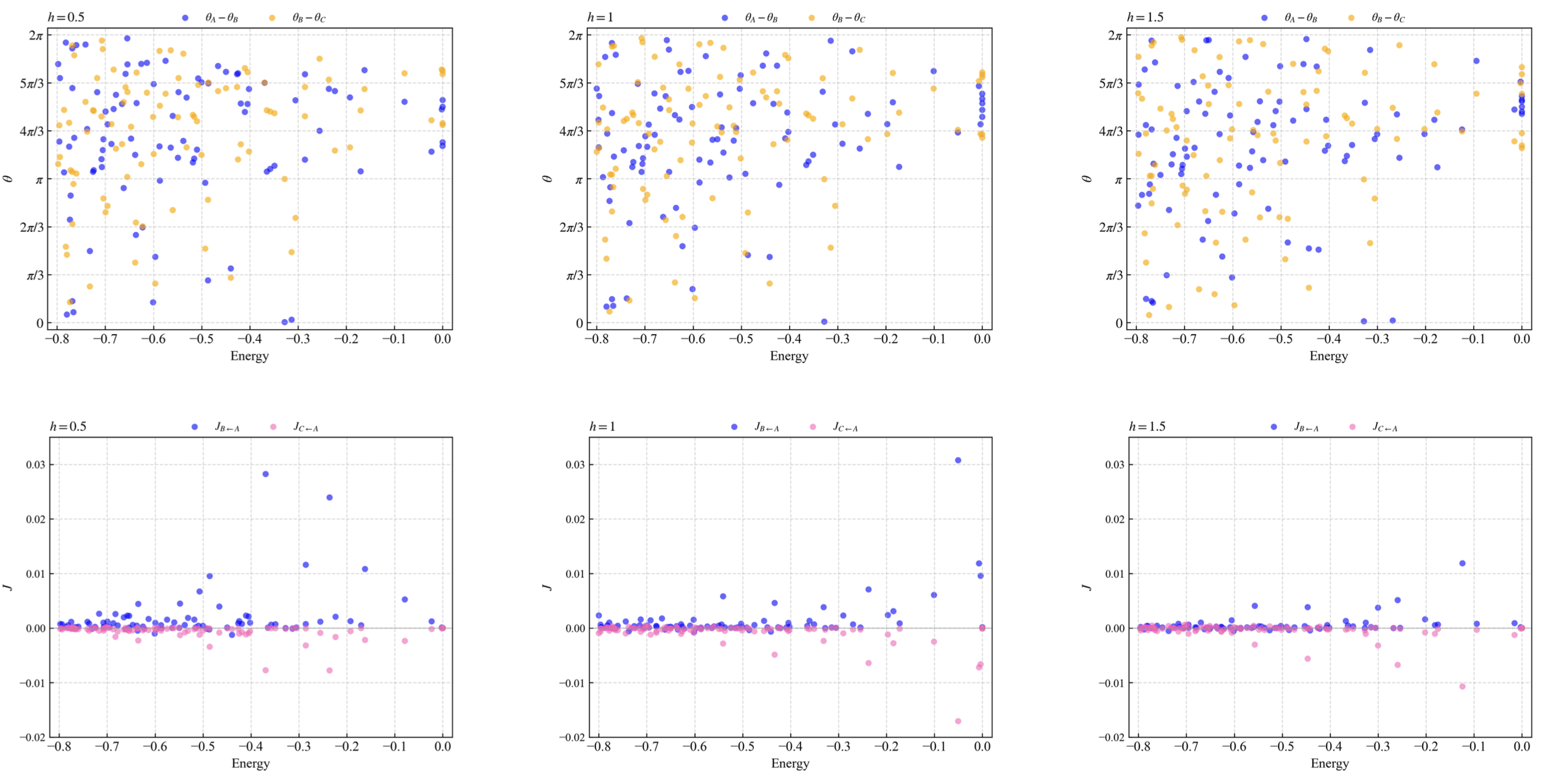}
    \caption{The phase difference and the currents with respect to energy  (for $L=25)$. The system is fixed at $h=0.5$, $1$, and $1.5$ (left,middle, right columns respectively), and the x-axis represents the selected range of negative energy. Each datapoint of a given color is a single state. 
    Top row: $\theta_A-\theta_B$ and $\theta_B-\theta_C$ for low-energy occupied states. Due to the absence of perfect rotational symmetry, the values of $\theta_A-\theta_B$ and $\theta_B-\theta_C$ no longer coincide. But it is clearly observed that in the low-energy region on the right side of each figure (e.g. [-0.1, 0]), the phase differences are concentrated around $4\pi/3$ and $5\pi/3$, which is consistent with the behavior of the toy model. 
    Bottom row: The currents on NN hopping and NNN hopping. 
    Large currents are observed from the low energy states near the transition. 
    }
        \label{fig:25}
\end{figure*}

Fig.~\ref{fig:toy} and Fig.~\ref{fig:25} show the correspondence between the energy of  low energy states and their phase difference between neighboring sites, for the six-site toy model and for the full system of $L= 25$ respectively. These results reveal that even as the system becomes more complicated, the low energy states still follow a similar pattern, where the phase difference $\theta$ tends to cluster around $4\pi/3$ and $5\pi/3$. %

\bibliography{references}

\begin{thebibliography}{43}%
\makeatletter
\providecommand \@ifxundefined [1]{%
 \@ifx{#1\undefined}
}%
\providecommand \@ifnum [1]{%
 \ifnum #1\expandafter \@firstoftwo
 \else \expandafter \@secondoftwo
 \fi
}%
\providecommand \@ifx [1]{%
 \ifx #1\expandafter \@firstoftwo
 \else \expandafter \@secondoftwo
 \fi
}%
\providecommand \natexlab [1]{#1}%
\providecommand \enquote  [1]{``#1''}%
\providecommand \bibnamefont  [1]{#1}%
\providecommand \bibfnamefont [1]{#1}%
\providecommand \citenamefont [1]{#1}%
\providecommand \href@noop [0]{\@secondoftwo}%
\providecommand \href [0]{\begingroup \@sanitize@url \@href}%
\providecommand \@href[1]{\@@startlink{#1}\@@href}%
\providecommand \@@href[1]{\endgroup#1\@@endlink}%
\providecommand \@sanitize@url [0]{\catcode `\\12\catcode `\$12\catcode `\&12\catcode `\#12\catcode `\^12\catcode `\_12\catcode `\%12\relax}%
\providecommand \@@startlink[1]{}%
\providecommand \@@endlink[0]{}%
\providecommand \url  [0]{\begingroup\@sanitize@url \@url }%
\providecommand \@url [1]{\endgroup\@href {#1}{\urlprefix }}%
\providecommand \urlprefix  [0]{URL }%
\providecommand \Eprint [0]{\href }%
\providecommand \doibase [0]{https://doi.org/}%
\providecommand \selectlanguage [0]{\@gobble}%
\providecommand \bibinfo  [0]{\@secondoftwo}%
\providecommand \bibfield  [0]{\@secondoftwo}%
\providecommand \translation [1]{[#1]}%
\providecommand \BibitemOpen [0]{}%
\providecommand \bibitemStop [0]{}%
\providecommand \bibitemNoStop [0]{.\EOS\space}%
\providecommand \EOS [0]{\spacefactor3000\relax}%
\providecommand \BibitemShut  [1]{\csname bibitem#1\endcsname}%
\let\auto@bib@innerbib\@empty
\bibitem [{\citenamefont {Haldane}(1988)}]{haldane_model_1988}%
  \BibitemOpen
  \bibfield  {author} {\bibinfo {author} {\bibfnamefont {F.~D.~M.}\ \bibnamefont {Haldane}},\ }\bibfield  {title} {\bibinfo {title} {Model for a {{Quantum Hall Effect}} without {{Landau Levels}}: {{Condensed-Matter Realization}} of the "{{Parity Anomaly}}"},\ }\href {https://doi.org/10.1103/PhysRevLett.61.2015} {\bibfield  {journal} {\bibinfo  {journal} {Physical Review Letters}\ }\textbf {\bibinfo {volume} {61}},\ \bibinfo {pages} {2015} (\bibinfo {year} {1988})}\BibitemShut {NoStop}%
\bibitem [{\citenamefont {Wehling}\ \emph {et~al.}(2007)\citenamefont {Wehling}, \citenamefont {Balatsky}, \citenamefont {Katsnelson}, \citenamefont {Lichtenstein}, \citenamefont {Scharnberg},\ and\ \citenamefont {Wiesendanger}}]{wehling_local_2007}%
  \BibitemOpen
  \bibfield  {author} {\bibinfo {author} {\bibfnamefont {T.~O.}\ \bibnamefont {Wehling}}, \bibinfo {author} {\bibfnamefont {A.~V.}\ \bibnamefont {Balatsky}}, \bibinfo {author} {\bibfnamefont {M.~I.}\ \bibnamefont {Katsnelson}}, \bibinfo {author} {\bibfnamefont {A.~I.}\ \bibnamefont {Lichtenstein}}, \bibinfo {author} {\bibfnamefont {K.}~\bibnamefont {Scharnberg}},\ and\ \bibinfo {author} {\bibfnamefont {R.}~\bibnamefont {Wiesendanger}},\ }\bibfield  {title} {\bibinfo {title} {Local electronic signatures of impurity states in graphene},\ }\href {https://doi.org/10.1103/PhysRevB.75.125425} {\bibfield  {journal} {\bibinfo  {journal} {Physical Review B}\ }\textbf {\bibinfo {volume} {75}},\ \bibinfo {pages} {125425} (\bibinfo {year} {2007})}\BibitemShut {NoStop}%
\bibitem [{\citenamefont {Yazyev}\ and\ \citenamefont {Helm}(2007)}]{yazyev_defectinduced_2007}%
  \BibitemOpen
  \bibfield  {author} {\bibinfo {author} {\bibfnamefont {O.~V.}\ \bibnamefont {Yazyev}}\ and\ \bibinfo {author} {\bibfnamefont {L.}~\bibnamefont {Helm}},\ }\bibfield  {title} {\bibinfo {title} {Defect-induced magnetism in graphene},\ }\href {https://doi.org/10.1103/PhysRevB.75.125408} {\bibfield  {journal} {\bibinfo  {journal} {Physical Review B}\ }\textbf {\bibinfo {volume} {75}},\ \bibinfo {pages} {125408} (\bibinfo {year} {2007})}\BibitemShut {NoStop}%
\bibitem [{\citenamefont {Ugeda}\ \emph {et~al.}(2010)\citenamefont {Ugeda}, \citenamefont {Brihuega}, \citenamefont {Guinea},\ and\ \citenamefont {{G{\'o}mez-Rodr{\'\i}guez}}}]{ugeda_missing_2010}%
  \BibitemOpen
  \bibfield  {author} {\bibinfo {author} {\bibfnamefont {M.~M.}\ \bibnamefont {Ugeda}}, \bibinfo {author} {\bibfnamefont {I.}~\bibnamefont {Brihuega}}, \bibinfo {author} {\bibfnamefont {F.}~\bibnamefont {Guinea}},\ and\ \bibinfo {author} {\bibfnamefont {J.~M.}\ \bibnamefont {{G{\'o}mez-Rodr{\'\i}guez}}},\ }\bibfield  {title} {\bibinfo {title} {Missing atom as a source of carbon magnetism},\ }\href {https://doi.org/10.1103/PhysRevLett.104.096804} {\bibfield  {journal} {\bibinfo  {journal} {Physical Review Letters}\ }\textbf {\bibinfo {volume} {104}},\ \bibinfo {pages} {096804} (\bibinfo {year} {2010})}\BibitemShut {NoStop}%
\bibitem [{\citenamefont {Lee}\ \emph {et~al.}(2005)\citenamefont {Lee}, \citenamefont {Son}, \citenamefont {Park}, \citenamefont {Han},\ and\ \citenamefont {Yu}}]{lee_magnetic_2005}%
  \BibitemOpen
  \bibfield  {author} {\bibinfo {author} {\bibfnamefont {H.}~\bibnamefont {Lee}}, \bibinfo {author} {\bibfnamefont {Y.-W.}\ \bibnamefont {Son}}, \bibinfo {author} {\bibfnamefont {N.}~\bibnamefont {Park}}, \bibinfo {author} {\bibfnamefont {S.}~\bibnamefont {Han}},\ and\ \bibinfo {author} {\bibfnamefont {J.}~\bibnamefont {Yu}},\ }\bibfield  {title} {\bibinfo {title} {Magnetic ordering at the edges of graphitic fragments: {{Magnetic}} tail interactions between the edge-localized states},\ }\href {https://doi.org/10.1103/PhysRevB.72.174431} {\bibfield  {journal} {\bibinfo  {journal} {Physical Review B}\ }\textbf {\bibinfo {volume} {72}},\ \bibinfo {pages} {174431} (\bibinfo {year} {2005})}\BibitemShut {NoStop}%
\bibitem [{\citenamefont {Kimchi}\ \emph {et~al.}(2013)\citenamefont {Kimchi}, \citenamefont {Parameswaran}, \citenamefont {Turner}, \citenamefont {Wang},\ and\ \citenamefont {Vishwanath}}]{kimchi_featureless_2013}%
  \BibitemOpen
  \bibfield  {author} {\bibinfo {author} {\bibfnamefont {I.}~\bibnamefont {Kimchi}}, \bibinfo {author} {\bibfnamefont {S.~A.}\ \bibnamefont {Parameswaran}}, \bibinfo {author} {\bibfnamefont {A.~M.}\ \bibnamefont {Turner}}, \bibinfo {author} {\bibfnamefont {F.}~\bibnamefont {Wang}},\ and\ \bibinfo {author} {\bibfnamefont {A.}~\bibnamefont {Vishwanath}},\ }\bibfield  {title} {\bibinfo {title} {Featureless and nonfractionalized {{Mott}} insulators on the honeycomb lattice at 1/2 site filling},\ }\href {https://doi.org/10.1073/pnas.1307245110} {\bibfield  {journal} {\bibinfo  {journal} {Proceedings of the National Academy of Sciences}\ }\textbf {\bibinfo {volume} {110}},\ \bibinfo {pages} {16378} (\bibinfo {year} {2013})}\BibitemShut {NoStop}%
\bibitem [{\citenamefont {Neehus}\ \emph {et~al.}(2025)\citenamefont {Neehus}, \citenamefont {Pollmann},\ and\ \citenamefont {Knolle}}]{neehus_genuine_2025}%
  \BibitemOpen
  \bibfield  {author} {\bibinfo {author} {\bibfnamefont {A.}~\bibnamefont {Neehus}}, \bibinfo {author} {\bibfnamefont {F.}~\bibnamefont {Pollmann}},\ and\ \bibinfo {author} {\bibfnamefont {J.}~\bibnamefont {Knolle}},\ }\bibfield  {title} {\bibinfo {title} {Genuine topological anderson insulator from impurity induced chirality reversal},\ }\href {https://doi.org/10.1103/7p8y-2mp6} {\bibfield  {journal} {\bibinfo  {journal} {Physical Review Letters}\ }\textbf {\bibinfo {volume} {135}},\ \bibinfo {pages} {126604} (\bibinfo {year} {2025})}\BibitemShut {NoStop}%
\bibitem [{\citenamefont {Jha}\ \emph {et~al.}(2017)\citenamefont {Jha}, \citenamefont {Rani},\ and\ \citenamefont {Ganesh}}]{jha_impurityinduced_2017}%
  \BibitemOpen
  \bibfield  {author} {\bibinfo {author} {\bibfnamefont {V.~B.}\ \bibnamefont {Jha}}, \bibinfo {author} {\bibfnamefont {G.}~\bibnamefont {Rani}},\ and\ \bibinfo {author} {\bibfnamefont {R.}~\bibnamefont {Ganesh}},\ }\bibfield  {title} {\bibinfo {title} {Impurity-induced current in a {{Chern}} insulator},\ }\href {https://doi.org/10.1103/PhysRevB.95.115434} {\bibfield  {journal} {\bibinfo  {journal} {Physical Review B}\ }\textbf {\bibinfo {volume} {95}},\ \bibinfo {pages} {115434} (\bibinfo {year} {2017})}\BibitemShut {NoStop}%
\bibitem [{\citenamefont {Karmakar}\ and\ \citenamefont {Ganesh}(2021)}]{karmakar_disorderinduced_2021}%
  \BibitemOpen
  \bibfield  {author} {\bibinfo {author} {\bibfnamefont {M.}~\bibnamefont {Karmakar}}\ and\ \bibinfo {author} {\bibfnamefont {R.}~\bibnamefont {Ganesh}},\ }\bibfield  {title} {\bibinfo {title} {Disorder-induced currents as signatures of chiral superconductivity},\ }\href {https://doi.org/10.1103/PhysRevB.104.094505} {\bibfield  {journal} {\bibinfo  {journal} {Physical Review B}\ }\textbf {\bibinfo {volume} {104}},\ \bibinfo {pages} {094505} (\bibinfo {year} {2021})}\BibitemShut {NoStop}%
\bibitem [{\citenamefont {Gonz{\'a}lez}\ and\ \citenamefont {{Fern{\'a}ndez-Rossier}}(2012)}]{gonzalez_impurity_2012}%
  \BibitemOpen
  \bibfield  {author} {\bibinfo {author} {\bibfnamefont {J.~W.}\ \bibnamefont {Gonz{\'a}lez}}\ and\ \bibinfo {author} {\bibfnamefont {J.}~\bibnamefont {{Fern{\'a}ndez-Rossier}}},\ }\bibfield  {title} {\bibinfo {title} {Impurity states in the quantum spin {{Hall}} phase in graphene},\ }\href {https://doi.org/10.1103/PhysRevB.86.115327} {\bibfield  {journal} {\bibinfo  {journal} {Physical Review B}\ }\textbf {\bibinfo {volume} {86}},\ \bibinfo {pages} {115327} (\bibinfo {year} {2012})}\BibitemShut {NoStop}%
\bibitem [{\citenamefont {Michel}\ \emph {et~al.}(2024)\citenamefont {Michel}, \citenamefont {F{\"u}nfhaus}, \citenamefont {Quade}, \citenamefont {Valent{\'\i}},\ and\ \citenamefont {Potthoff}}]{michel_bound_2024}%
  \BibitemOpen
  \bibfield  {author} {\bibinfo {author} {\bibfnamefont {S.}~\bibnamefont {Michel}}, \bibinfo {author} {\bibfnamefont {A.}~\bibnamefont {F{\"u}nfhaus}}, \bibinfo {author} {\bibfnamefont {R.}~\bibnamefont {Quade}}, \bibinfo {author} {\bibfnamefont {R.}~\bibnamefont {Valent{\'\i}}},\ and\ \bibinfo {author} {\bibfnamefont {M.}~\bibnamefont {Potthoff}},\ }\bibfield  {title} {\bibinfo {title} {Bound states and local topological phase diagram of classical impurity spins coupled to a {{Chern}} insulator},\ }\href {https://doi.org/10.1103/PhysRevB.109.155116} {\bibfield  {journal} {\bibinfo  {journal} {Physical Review B}\ }\textbf {\bibinfo {volume} {109}},\ \bibinfo {pages} {155116} (\bibinfo {year} {2024})}\BibitemShut {NoStop}%
\bibitem [{\citenamefont {Li}\ \emph {et~al.}(2009)\citenamefont {Li}, \citenamefont {Chu}, \citenamefont {Jain},\ and\ \citenamefont {Shen}}]{li_topological_2009}%
  \BibitemOpen
  \bibfield  {author} {\bibinfo {author} {\bibfnamefont {J.}~\bibnamefont {Li}}, \bibinfo {author} {\bibfnamefont {R.-L.}\ \bibnamefont {Chu}}, \bibinfo {author} {\bibfnamefont {J.~K.}\ \bibnamefont {Jain}},\ and\ \bibinfo {author} {\bibfnamefont {S.-Q.}\ \bibnamefont {Shen}},\ }\bibfield  {title} {\bibinfo {title} {Topological anderson insulator},\ }\href {https://doi.org/10.1103/PhysRevLett.102.136806} {\bibfield  {journal} {\bibinfo  {journal} {Physical Review Letters}\ }\textbf {\bibinfo {volume} {102}},\ \bibinfo {pages} {136806} (\bibinfo {year} {2009})}\BibitemShut {NoStop}%
\bibitem [{\citenamefont {Groth}\ \emph {et~al.}(2009)\citenamefont {Groth}, \citenamefont {Wimmer}, \citenamefont {Akhmerov}, \citenamefont {Tworzyd{\l}o},\ and\ \citenamefont {Beenakker}}]{groth_theory_2009}%
  \BibitemOpen
  \bibfield  {author} {\bibinfo {author} {\bibfnamefont {C.~W.}\ \bibnamefont {Groth}}, \bibinfo {author} {\bibfnamefont {M.}~\bibnamefont {Wimmer}}, \bibinfo {author} {\bibfnamefont {A.~R.}\ \bibnamefont {Akhmerov}}, \bibinfo {author} {\bibfnamefont {J.}~\bibnamefont {Tworzyd{\l}o}},\ and\ \bibinfo {author} {\bibfnamefont {C.~W.~J.}\ \bibnamefont {Beenakker}},\ }\bibfield  {title} {\bibinfo {title} {Theory of the topological anderson insulator},\ }\href {https://doi.org/10.1103/PhysRevLett.103.196805} {\bibfield  {journal} {\bibinfo  {journal} {Physical Review Letters}\ }\textbf {\bibinfo {volume} {103}},\ \bibinfo {pages} {196805} (\bibinfo {year} {2009})}\BibitemShut {NoStop}%
\bibitem [{\citenamefont {Meier}\ \emph {et~al.}(2018)\citenamefont {Meier}, \citenamefont {An}, \citenamefont {Dauphin}, \citenamefont {Maffei}, \citenamefont {Massignan}, \citenamefont {Hughes},\ and\ \citenamefont {Gadway}}]{meier_observation_2018}%
  \BibitemOpen
  \bibfield  {author} {\bibinfo {author} {\bibfnamefont {E.~J.}\ \bibnamefont {Meier}}, \bibinfo {author} {\bibfnamefont {F.~A.}\ \bibnamefont {An}}, \bibinfo {author} {\bibfnamefont {A.}~\bibnamefont {Dauphin}}, \bibinfo {author} {\bibfnamefont {M.}~\bibnamefont {Maffei}}, \bibinfo {author} {\bibfnamefont {P.}~\bibnamefont {Massignan}}, \bibinfo {author} {\bibfnamefont {T.~L.}\ \bibnamefont {Hughes}},\ and\ \bibinfo {author} {\bibfnamefont {B.}~\bibnamefont {Gadway}},\ }\bibfield  {title} {\bibinfo {title} {Observation of the topological {{Anderson}} insulator in disordered atomic wires},\ }\href {https://doi.org/10.1126/science.aat3406} {\bibfield  {journal} {\bibinfo  {journal} {Science}\ }\textbf {\bibinfo {volume} {362}},\ \bibinfo {pages} {929} (\bibinfo {year} {2018})}\BibitemShut {NoStop}%
\bibitem [{\citenamefont {Ryu}\ \emph {et~al.}(2010)\citenamefont {Ryu}, \citenamefont {Schnyder}, \citenamefont {Furusaki},\ and\ \citenamefont {Ludwig}}]{ryu_topological_2010}%
  \BibitemOpen
  \bibfield  {author} {\bibinfo {author} {\bibfnamefont {S.}~\bibnamefont {Ryu}}, \bibinfo {author} {\bibfnamefont {A.~P.}\ \bibnamefont {Schnyder}}, \bibinfo {author} {\bibfnamefont {A.}~\bibnamefont {Furusaki}},\ and\ \bibinfo {author} {\bibfnamefont {A.~W.~W.}\ \bibnamefont {Ludwig}},\ }\bibfield  {title} {\bibinfo {title} {Topological insulators and superconductors: Tenfold way and dimensional hierarchy},\ }\href {https://doi.org/10.1088/1367-2630/12/6/065010} {\bibfield  {journal} {\bibinfo  {journal} {New Journal of Physics}\ }\textbf {\bibinfo {volume} {12}},\ \bibinfo {pages} {065010} (\bibinfo {year} {2010})}\BibitemShut {NoStop}%
\bibitem [{\citenamefont {Kitaev}(2009)}]{kitaev_periodic_2009}%
  \BibitemOpen
  \bibfield  {author} {\bibinfo {author} {\bibfnamefont {A.}~\bibnamefont {Kitaev}},\ }\bibfield  {title} {\bibinfo {title} {Periodic table for topological insulators and superconductors},\ }\href {https://doi.org/10.1063/1.3149495} {\bibfield  {journal} {\bibinfo  {journal} {AIP Conference Proceedings}\ }\textbf {\bibinfo {volume} {1134}},\ \bibinfo {pages} {22} (\bibinfo {year} {2009})}\BibitemShut {NoStop}%
\bibitem [{\citenamefont {Ludwig}\ \emph {et~al.}(1994)\citenamefont {Ludwig}, \citenamefont {Fisher}, \citenamefont {Shankar},\ and\ \citenamefont {Grinstein}}]{ludwig_integer_1994}%
  \BibitemOpen
  \bibfield  {author} {\bibinfo {author} {\bibfnamefont {A.~W.~W.}\ \bibnamefont {Ludwig}}, \bibinfo {author} {\bibfnamefont {M.~P.~A.}\ \bibnamefont {Fisher}}, \bibinfo {author} {\bibfnamefont {R.}~\bibnamefont {Shankar}},\ and\ \bibinfo {author} {\bibfnamefont {G.}~\bibnamefont {Grinstein}},\ }\bibfield  {title} {\bibinfo {title} {Integer quantum {{Hall}} transition: {{An}} alternative approach and exact results},\ }\href {https://doi.org/10.1103/PhysRevB.50.7526} {\bibfield  {journal} {\bibinfo  {journal} {Physical Review B}\ }\textbf {\bibinfo {volume} {50}},\ \bibinfo {pages} {7526} (\bibinfo {year} {1994})}\BibitemShut {NoStop}%
\bibitem [{\citenamefont {Cho}\ and\ \citenamefont {Fisher}(1997)}]{cho_criticality_1997}%
  \BibitemOpen
  \bibfield  {author} {\bibinfo {author} {\bibfnamefont {S.}~\bibnamefont {Cho}}\ and\ \bibinfo {author} {\bibfnamefont {M.~P.~A.}\ \bibnamefont {Fisher}},\ }\bibfield  {title} {\bibinfo {title} {Criticality in the two-dimensional random-bond {{Ising}} model},\ }\href {https://doi.org/10.1103/PhysRevB.55.1025} {\bibfield  {journal} {\bibinfo  {journal} {Physical Review B}\ }\textbf {\bibinfo {volume} {55}},\ \bibinfo {pages} {1025} (\bibinfo {year} {1997})}\BibitemShut {NoStop}%
\bibitem [{\citenamefont {Bocquet}\ \emph {et~al.}(2000)\citenamefont {Bocquet}, \citenamefont {Serban},\ and\ \citenamefont {Zirnbauer}}]{bocquet_disordered_2000}%
  \BibitemOpen
  \bibfield  {author} {\bibinfo {author} {\bibfnamefont {M.}~\bibnamefont {Bocquet}}, \bibinfo {author} {\bibfnamefont {D.}~\bibnamefont {Serban}},\ and\ \bibinfo {author} {\bibfnamefont {M.~R.}\ \bibnamefont {Zirnbauer}},\ }\bibfield  {title} {\bibinfo {title} {Disordered 2d quasiparticles in class {{D}}: {{Dirac}} fermions with random mass, and dirty superconductors},\ }\href {https://doi.org/10.1016/S0550-3213(00)00208-X} {\bibfield  {journal} {\bibinfo  {journal} {Nuclear Physics B}\ }\textbf {\bibinfo {volume} {578}},\ \bibinfo {pages} {628} (\bibinfo {year} {2000})}\BibitemShut {NoStop}%
\bibitem [{\citenamefont {Merz}\ and\ \citenamefont {Chalker}(2002)}]{merz_twodimensional_2002}%
  \BibitemOpen
  \bibfield  {author} {\bibinfo {author} {\bibfnamefont {F.}~\bibnamefont {Merz}}\ and\ \bibinfo {author} {\bibfnamefont {J.~T.}\ \bibnamefont {Chalker}},\ }\bibfield  {title} {\bibinfo {title} {Two-dimensional random-bond {{Ising}} model, free fermions, and the network model},\ }\href {https://doi.org/10.1103/PhysRevB.65.054425} {\bibfield  {journal} {\bibinfo  {journal} {Physical Review B}\ }\textbf {\bibinfo {volume} {65}},\ \bibinfo {pages} {054425} (\bibinfo {year} {2002})}\BibitemShut {NoStop}%
\bibitem [{\citenamefont {Gruzberg}\ \emph {et~al.}(2001)\citenamefont {Gruzberg}, \citenamefont {Read},\ and\ \citenamefont {Ludwig}}]{gruzberg_randombond_2001}%
  \BibitemOpen
  \bibfield  {author} {\bibinfo {author} {\bibfnamefont {I.~A.}\ \bibnamefont {Gruzberg}}, \bibinfo {author} {\bibfnamefont {N.}~\bibnamefont {Read}},\ and\ \bibinfo {author} {\bibfnamefont {A.~W.~W.}\ \bibnamefont {Ludwig}},\ }\bibfield  {title} {\bibinfo {title} {Random-bond {{Ising}} model in two dimensions: {{The Nishimori}} line and supersymmetry},\ }\href {https://doi.org/10.1103/PhysRevB.63.104422} {\bibfield  {journal} {\bibinfo  {journal} {Physical Review B}\ }\textbf {\bibinfo {volume} {63}},\ \bibinfo {pages} {104422} (\bibinfo {year} {2001})}\BibitemShut {NoStop}%
\bibitem [{\citenamefont {Chalker}\ \emph {et~al.}(2001)\citenamefont {Chalker}, \citenamefont {Read}, \citenamefont {Kagalovsky}, \citenamefont {Horovitz}, \citenamefont {Avishai},\ and\ \citenamefont {Ludwig}}]{chalker_thermal_2001}%
  \BibitemOpen
  \bibfield  {author} {\bibinfo {author} {\bibfnamefont {J.~T.}\ \bibnamefont {Chalker}}, \bibinfo {author} {\bibfnamefont {N.}~\bibnamefont {Read}}, \bibinfo {author} {\bibfnamefont {V.}~\bibnamefont {Kagalovsky}}, \bibinfo {author} {\bibfnamefont {B.}~\bibnamefont {Horovitz}}, \bibinfo {author} {\bibfnamefont {Y.}~\bibnamefont {Avishai}},\ and\ \bibinfo {author} {\bibfnamefont {A.~W.~W.}\ \bibnamefont {Ludwig}},\ }\bibfield  {title} {\bibinfo {title} {Thermal metal in network models of a disordered two-dimensional superconductor},\ }\href {https://doi.org/10.1103/PhysRevB.65.012506} {\bibfield  {journal} {\bibinfo  {journal} {Physical Review B}\ }\textbf {\bibinfo {volume} {65}},\ \bibinfo {pages} {012506} (\bibinfo {year} {2001})}\BibitemShut {NoStop}%
\bibitem [{\citenamefont {Medvedyeva}\ \emph {et~al.}(2010)\citenamefont {Medvedyeva}, \citenamefont {Tworzyd{\l}o},\ and\ \citenamefont {Beenakker}}]{medvedyeva_effective_2010}%
  \BibitemOpen
  \bibfield  {author} {\bibinfo {author} {\bibfnamefont {M.~V.}\ \bibnamefont {Medvedyeva}}, \bibinfo {author} {\bibfnamefont {J.}~\bibnamefont {Tworzyd{\l}o}},\ and\ \bibinfo {author} {\bibfnamefont {C.~W.~J.}\ \bibnamefont {Beenakker}},\ }\bibfield  {title} {\bibinfo {title} {Effective mass and tricritical point for lattice fermions localized by a random mass},\ }\href {https://doi.org/10.1103/PhysRevB.81.214203} {\bibfield  {journal} {\bibinfo  {journal} {Physical Review B}\ }\textbf {\bibinfo {volume} {81}},\ \bibinfo {pages} {214203} (\bibinfo {year} {2010})}\BibitemShut {NoStop}%
\bibitem [{\citenamefont {Mildenberger}\ \emph {et~al.}(2006)\citenamefont {Mildenberger}, \citenamefont {Evers}, \citenamefont {Narayanan}, \citenamefont {Mirlin},\ and\ \citenamefont {Damle}}]{mildenberger_griffiths_2006}%
  \BibitemOpen
  \bibfield  {author} {\bibinfo {author} {\bibfnamefont {A.}~\bibnamefont {Mildenberger}}, \bibinfo {author} {\bibfnamefont {F.}~\bibnamefont {Evers}}, \bibinfo {author} {\bibfnamefont {R.}~\bibnamefont {Narayanan}}, \bibinfo {author} {\bibfnamefont {A.~D.}\ \bibnamefont {Mirlin}},\ and\ \bibinfo {author} {\bibfnamefont {K.}~\bibnamefont {Damle}},\ }\bibfield  {title} {\bibinfo {title} {Griffiths phase in the thermal quantum {{Hall}} effect},\ }\href {https://doi.org/10.1103/PhysRevB.73.121301} {\bibfield  {journal} {\bibinfo  {journal} {Physical Review B}\ }\textbf {\bibinfo {volume} {73}},\ \bibinfo {pages} {121301} (\bibinfo {year} {2006})}\BibitemShut {NoStop}%
\bibitem [{\citenamefont {Huang}\ \emph {et~al.}(2014)\citenamefont {Huang}, \citenamefont {Taylor},\ and\ \citenamefont {Kallin}}]{huang_vanishing_2014}%
  \BibitemOpen
  \bibfield  {author} {\bibinfo {author} {\bibfnamefont {W.}~\bibnamefont {Huang}}, \bibinfo {author} {\bibfnamefont {E.}~\bibnamefont {Taylor}},\ and\ \bibinfo {author} {\bibfnamefont {C.}~\bibnamefont {Kallin}},\ }\bibfield  {title} {\bibinfo {title} {Vanishing edge currents in non-{{p}}-wave topological chiral superconductors},\ }\href {https://doi.org/10.1103/PhysRevB.90.224519} {\bibfield  {journal} {\bibinfo  {journal} {Physical Review B}\ }\textbf {\bibinfo {volume} {90}},\ \bibinfo {pages} {224519} (\bibinfo {year} {2014})}\BibitemShut {NoStop}%
\bibitem [{\citenamefont {Bianco}\ and\ \citenamefont {Resta}(2011)}]{bianco_mapping_2011}%
  \BibitemOpen
  \bibfield  {author} {\bibinfo {author} {\bibfnamefont {R.}~\bibnamefont {Bianco}}\ and\ \bibinfo {author} {\bibfnamefont {R.}~\bibnamefont {Resta}},\ }\bibfield  {title} {\bibinfo {title} {Mapping topological order in coordinate space},\ }\href {https://doi.org/10.1103/PhysRevB.84.241106} {\bibfield  {journal} {\bibinfo  {journal} {Physical Review B}\ }\textbf {\bibinfo {volume} {84}},\ \bibinfo {pages} {241106} (\bibinfo {year} {2011})}\BibitemShut {NoStop}%
\bibitem [{\citenamefont {Bianco}\ and\ \citenamefont {Resta}(2013)}]{bianco_orbital_2013}%
  \BibitemOpen
  \bibfield  {author} {\bibinfo {author} {\bibfnamefont {R.}~\bibnamefont {Bianco}}\ and\ \bibinfo {author} {\bibfnamefont {R.}~\bibnamefont {Resta}},\ }\bibfield  {title} {\bibinfo {title} {Orbital {{Magnetization}} as a {{Local Property}}},\ }\href {https://doi.org/10.1103/PhysRevLett.110.087202} {\bibfield  {journal} {\bibinfo  {journal} {Physical Review Letters}\ }\textbf {\bibinfo {volume} {110}},\ \bibinfo {pages} {087202} (\bibinfo {year} {2013})}\BibitemShut {NoStop}%
\bibitem [{\citenamefont {Ceresoli}\ \emph {et~al.}(2006)\citenamefont {Ceresoli}, \citenamefont {Thonhauser}, \citenamefont {Vanderbilt},\ and\ \citenamefont {Resta}}]{ceresoli_orbital_2006}%
  \BibitemOpen
  \bibfield  {author} {\bibinfo {author} {\bibfnamefont {D.}~\bibnamefont {Ceresoli}}, \bibinfo {author} {\bibfnamefont {T.}~\bibnamefont {Thonhauser}}, \bibinfo {author} {\bibfnamefont {D.}~\bibnamefont {Vanderbilt}},\ and\ \bibinfo {author} {\bibfnamefont {R.}~\bibnamefont {Resta}},\ }\bibfield  {title} {\bibinfo {title} {Orbital magnetization in crystalline solids: {{Multi-band}} insulators, {{Chern}} insulators, and metals},\ }\href {https://doi.org/10.1103/PhysRevB.74.024408} {\bibfield  {journal} {\bibinfo  {journal} {Physical Review B}\ }\textbf {\bibinfo {volume} {74}},\ \bibinfo {pages} {024408} (\bibinfo {year} {2006})}\BibitemShut {NoStop}%
\bibitem [{\citenamefont {Jiang}\ \emph {et~al.}(2012)\citenamefont {Jiang}, \citenamefont {Qiao}, \citenamefont {Liu}, \citenamefont {Shi},\ and\ \citenamefont {Niu}}]{jiang_stabilizing_2012}%
  \BibitemOpen
  \bibfield  {author} {\bibinfo {author} {\bibfnamefont {H.}~\bibnamefont {Jiang}}, \bibinfo {author} {\bibfnamefont {Z.}~\bibnamefont {Qiao}}, \bibinfo {author} {\bibfnamefont {H.}~\bibnamefont {Liu}}, \bibinfo {author} {\bibfnamefont {J.}~\bibnamefont {Shi}},\ and\ \bibinfo {author} {\bibfnamefont {Q.}~\bibnamefont {Niu}},\ }\bibfield  {title} {\bibinfo {title} {Stabilizing topological phases in graphene via random adsorption},\ }\href {https://doi.org/10.1103/PhysRevLett.109.116803} {\bibfield  {journal} {\bibinfo  {journal} {Physical Review Letters}\ }\textbf {\bibinfo {volume} {109}},\ \bibinfo {pages} {116803} (\bibinfo {year} {2012})}\BibitemShut {NoStop}%
\bibitem [{\citenamefont {Kot}\ \emph {et~al.}(2020)\citenamefont {Kot}, \citenamefont {Parnell}, \citenamefont {Habibian}, \citenamefont {Stra{\ss}er}, \citenamefont {Ostrovsky},\ and\ \citenamefont {Ast}}]{kot_band_2020}%
  \BibitemOpen
  \bibfield  {author} {\bibinfo {author} {\bibfnamefont {P.}~\bibnamefont {Kot}}, \bibinfo {author} {\bibfnamefont {J.}~\bibnamefont {Parnell}}, \bibinfo {author} {\bibfnamefont {S.}~\bibnamefont {Habibian}}, \bibinfo {author} {\bibfnamefont {C.}~\bibnamefont {Stra{\ss}er}}, \bibinfo {author} {\bibfnamefont {P.~M.}\ \bibnamefont {Ostrovsky}},\ and\ \bibinfo {author} {\bibfnamefont {C.~R.}\ \bibnamefont {Ast}},\ }\bibfield  {title} {\bibinfo {title} {Band dispersion of graphene with structural defects},\ }\href {https://doi.org/10.1103/PhysRevB.101.235116} {\bibfield  {journal} {\bibinfo  {journal} {Physical Review B}\ }\textbf {\bibinfo {volume} {101}},\ \bibinfo {pages} {235116} (\bibinfo {year} {2020})}\BibitemShut {NoStop}%
\bibitem [{\citenamefont {Settnes}\ \emph {et~al.}(2015)\citenamefont {Settnes}, \citenamefont {Power}, \citenamefont {Lin}, \citenamefont {Petersen},\ and\ \citenamefont {Jauho}}]{settnes_patched_2015}%
  \BibitemOpen
  \bibfield  {author} {\bibinfo {author} {\bibfnamefont {M.}~\bibnamefont {Settnes}}, \bibinfo {author} {\bibfnamefont {S.~R.}\ \bibnamefont {Power}}, \bibinfo {author} {\bibfnamefont {J.}~\bibnamefont {Lin}}, \bibinfo {author} {\bibfnamefont {D.~H.}\ \bibnamefont {Petersen}},\ and\ \bibinfo {author} {\bibfnamefont {A.-P.}\ \bibnamefont {Jauho}},\ }\bibfield  {title} {\bibinfo {title} {Patched {{Green}}'s function techniques for two-dimensional systems: {{Electronic}} behavior of bubbles and perforations in graphene},\ }\bibfield  {journal} {\bibinfo  {journal} {Physical Review B}\ }\textbf {\bibinfo {volume} {91}},\ \href {https://doi.org/10.1103/physrevb.91.125408} {10.1103/physrevb.91.125408} (\bibinfo {year} {2015})\BibitemShut {NoStop}%
\bibitem [{\citenamefont {Groth}\ \emph {et~al.}(2014)\citenamefont {Groth}, \citenamefont {Wimmer}, \citenamefont {Akhmerov},\ and\ \citenamefont {Waintal}}]{groth_kwant_2014}%
  \BibitemOpen
  \bibfield  {author} {\bibinfo {author} {\bibfnamefont {C.~W.}\ \bibnamefont {Groth}}, \bibinfo {author} {\bibfnamefont {M.}~\bibnamefont {Wimmer}}, \bibinfo {author} {\bibfnamefont {A.~R.}\ \bibnamefont {Akhmerov}},\ and\ \bibinfo {author} {\bibfnamefont {X.}~\bibnamefont {Waintal}},\ }\bibfield  {title} {\bibinfo {title} {Kwant: A software package for quantum transport},\ }\href {https://doi.org/10.1088/1367-2630/16/6/063065} {\bibfield  {journal} {\bibinfo  {journal} {New Journal of Physics}\ }\textbf {\bibinfo {volume} {16}},\ \bibinfo {pages} {063065} (\bibinfo {year} {2014})}\BibitemShut {NoStop}%
\bibitem [{\citenamefont {Ba{\`u}}\ and\ \citenamefont {Marrazzo}(2024)}]{bau_local_2024}%
  \BibitemOpen
  \bibfield  {author} {\bibinfo {author} {\bibfnamefont {N.}~\bibnamefont {Ba{\`u}}}\ and\ \bibinfo {author} {\bibfnamefont {A.}~\bibnamefont {Marrazzo}},\ }\bibfield  {title} {\bibinfo {title} {Local {{Chern Marker}} for {{Periodic Systems}}},\ }\href {https://doi.org/10.1103/PhysRevB.109.014206} {\bibfield  {journal} {\bibinfo  {journal} {Physical Review B}\ }\textbf {\bibinfo {volume} {109}},\ \bibinfo {pages} {014206} (\bibinfo {year} {2024})},\ \Eprint {https://arxiv.org/abs/2310.15783} {arXiv:2310.15783 [cond-mat]} \BibitemShut {NoStop}%
\bibitem [{\citenamefont {Slager}\ \emph {et~al.}(2015)\citenamefont {Slager}, \citenamefont {Rademaker}, \citenamefont {Zaanen},\ and\ \citenamefont {Balents}}]{slager_impuritybound_2015}%
  \BibitemOpen
  \bibfield  {author} {\bibinfo {author} {\bibfnamefont {R.-J.}\ \bibnamefont {Slager}}, \bibinfo {author} {\bibfnamefont {L.}~\bibnamefont {Rademaker}}, \bibinfo {author} {\bibfnamefont {J.}~\bibnamefont {Zaanen}},\ and\ \bibinfo {author} {\bibfnamefont {L.}~\bibnamefont {Balents}},\ }\bibfield  {title} {\bibinfo {title} {Impurity-bound states and {{Green}}'s function zeros as local signatures of topology},\ }\href {https://doi.org/10.1103/PhysRevB.92.085126} {\bibfield  {journal} {\bibinfo  {journal} {Physical Review B}\ }\textbf {\bibinfo {volume} {92}},\ \bibinfo {pages} {085126} (\bibinfo {year} {2015})}\BibitemShut {NoStop}%
\bibitem [{\citenamefont {Queiroz}\ \emph {et~al.}(2024)\citenamefont {Queiroz}, \citenamefont {Ilan}, \citenamefont {Song}, \citenamefont {Bernevig},\ and\ \citenamefont {Stern}}]{queiroz_ring_2024}%
  \BibitemOpen
  \bibfield  {author} {\bibinfo {author} {\bibfnamefont {R.}~\bibnamefont {Queiroz}}, \bibinfo {author} {\bibfnamefont {R.}~\bibnamefont {Ilan}}, \bibinfo {author} {\bibfnamefont {Z.}~\bibnamefont {Song}}, \bibinfo {author} {\bibfnamefont {B.~A.}\ \bibnamefont {Bernevig}},\ and\ \bibinfo {author} {\bibfnamefont {A.}~\bibnamefont {Stern}},\ }\href {https://doi.org/10.48550/arXiv.2406.03529} {\bibinfo {title} {Ring states in topological materials}} (\bibinfo {year} {2024}),\ \Eprint {https://arxiv.org/abs/2406.03529} {arXiv:2406.03529 [cond-mat]} \BibitemShut {NoStop}%
\bibitem [{\citenamefont {Kitaev}(2006)}]{kitaev_anyons_2006}%
  \BibitemOpen
  \bibfield  {author} {\bibinfo {author} {\bibfnamefont {A.}~\bibnamefont {Kitaev}},\ }\bibfield  {title} {\bibinfo {title} {Anyons in an exactly solved model and beyond},\ }\href {https://doi.org/10.1016/j.aop.2005.10.005} {\bibfield  {journal} {\bibinfo  {journal} {Annals of Physics}\ }\textbf {\bibinfo {volume} {321}},\ \bibinfo {pages} {2} (\bibinfo {year} {2006})}\BibitemShut {NoStop}%
\bibitem [{\citenamefont {Dhochak}\ \emph {et~al.}(2010)\citenamefont {Dhochak}, \citenamefont {Shankar},\ and\ \citenamefont {Tripathi}}]{dhochak_magnetic_2010}%
  \BibitemOpen
  \bibfield  {author} {\bibinfo {author} {\bibfnamefont {K.}~\bibnamefont {Dhochak}}, \bibinfo {author} {\bibfnamefont {R.}~\bibnamefont {Shankar}},\ and\ \bibinfo {author} {\bibfnamefont {V.}~\bibnamefont {Tripathi}},\ }\bibfield  {title} {\bibinfo {title} {Magnetic {{Impurities}} in the {{Honeycomb Kitaev Model}}},\ }\href {https://doi.org/10.1103/PhysRevLett.105.117201} {\bibfield  {journal} {\bibinfo  {journal} {Physical Review Letters}\ }\textbf {\bibinfo {volume} {105}},\ \bibinfo {pages} {117201} (\bibinfo {year} {2010})}\BibitemShut {NoStop}%
\bibitem [{\citenamefont {Keskiner}\ \emph {et~al.}(2025)\citenamefont {Keskiner}, \citenamefont {Oktel}, \citenamefont {Perkins},\ and\ \citenamefont {Erten}}]{keskiner_magnetic_2025}%
  \BibitemOpen
  \bibfield  {author} {\bibinfo {author} {\bibfnamefont {M.~A.}\ \bibnamefont {Keskiner}}, \bibinfo {author} {\bibfnamefont {M.~{\"O}.}\ \bibnamefont {Oktel}}, \bibinfo {author} {\bibfnamefont {N.~B.}\ \bibnamefont {Perkins}},\ and\ \bibinfo {author} {\bibfnamefont {O.}~\bibnamefont {Erten}},\ }\bibfield  {title} {\bibinfo {title} {Magnetic order through {{Kondo}} coupling to quantum spin liquids},\ }\href {https://doi.org/10.1016/j.mtquan.2025.100038} {\bibfield  {journal} {\bibinfo  {journal} {Materials Today Quantum}\ }\textbf {\bibinfo {volume} {6}},\ \bibinfo {pages} {100038} (\bibinfo {year} {2025})}\BibitemShut {NoStop}%
\bibitem [{\citenamefont {Du}\ \emph {et~al.}(2017)\citenamefont {Du}, \citenamefont {{van der Sar}}, \citenamefont {Zhou}, \citenamefont {Upadhyaya}, \citenamefont {Casola}, \citenamefont {Zhang}, \citenamefont {Onbasli}, \citenamefont {Ross}, \citenamefont {Walsworth}, \citenamefont {Tserkovnyak},\ and\ \citenamefont {Yacoby}}]{du_control_2017}%
  \BibitemOpen
  \bibfield  {author} {\bibinfo {author} {\bibfnamefont {C.}~\bibnamefont {Du}}, \bibinfo {author} {\bibfnamefont {T.}~\bibnamefont {{van der Sar}}}, \bibinfo {author} {\bibfnamefont {T.~X.}\ \bibnamefont {Zhou}}, \bibinfo {author} {\bibfnamefont {P.}~\bibnamefont {Upadhyaya}}, \bibinfo {author} {\bibfnamefont {F.}~\bibnamefont {Casola}}, \bibinfo {author} {\bibfnamefont {H.}~\bibnamefont {Zhang}}, \bibinfo {author} {\bibfnamefont {M.~C.}\ \bibnamefont {Onbasli}}, \bibinfo {author} {\bibfnamefont {C.~A.}\ \bibnamefont {Ross}}, \bibinfo {author} {\bibfnamefont {R.~L.}\ \bibnamefont {Walsworth}}, \bibinfo {author} {\bibfnamefont {Y.}~\bibnamefont {Tserkovnyak}},\ and\ \bibinfo {author} {\bibfnamefont {A.}~\bibnamefont {Yacoby}},\ }\bibfield  {title} {\bibinfo {title} {Control and local measurement of the spin chemical potential in a magnetic insulator},\ }\href {https://doi.org/10.1126/science.aak9611} {\bibfield  {journal} {\bibinfo  {journal} {Science}\ }\textbf {\bibinfo {volume} {357}},\ \bibinfo {pages}
  {195} (\bibinfo {year} {2017})}\BibitemShut {NoStop}%
\bibitem [{\citenamefont {Tetienne}\ \emph {et~al.}(2017)\citenamefont {Tetienne}, \citenamefont {Dontschuk}, \citenamefont {Broadway}, \citenamefont {Stacey}, \citenamefont {Simpson},\ and\ \citenamefont {Hollenberg}}]{tetienne_quantum_2017}%
  \BibitemOpen
  \bibfield  {author} {\bibinfo {author} {\bibfnamefont {J.-P.}\ \bibnamefont {Tetienne}}, \bibinfo {author} {\bibfnamefont {N.}~\bibnamefont {Dontschuk}}, \bibinfo {author} {\bibfnamefont {D.~A.}\ \bibnamefont {Broadway}}, \bibinfo {author} {\bibfnamefont {A.}~\bibnamefont {Stacey}}, \bibinfo {author} {\bibfnamefont {D.~A.}\ \bibnamefont {Simpson}},\ and\ \bibinfo {author} {\bibfnamefont {L.~C.~L.}\ \bibnamefont {Hollenberg}},\ }\bibfield  {title} {\bibinfo {title} {Quantum imaging of current flow in graphene},\ }\bibfield  {journal} {\bibinfo  {journal} {Science Advances}\ }\href {https://doi.org/10.1126/sciadv.1602429} {10.1126/sciadv.1602429} (\bibinfo {year} {2017})\BibitemShut {NoStop}%
\bibitem [{\citenamefont {McLaughlin}\ \emph {et~al.}(2023)\citenamefont {McLaughlin}, \citenamefont {Li}, \citenamefont {Brock}, \citenamefont {Zhang}, \citenamefont {Lu}, \citenamefont {Huang}, \citenamefont {Xiao}, \citenamefont {Zhou}, \citenamefont {Tserkovnyak}, \citenamefont {Fullerton}, \citenamefont {Wang},\ and\ \citenamefont {Du}}]{mclaughlin_local_2023}%
  \BibitemOpen
  \bibfield  {author} {\bibinfo {author} {\bibfnamefont {N.~J.}\ \bibnamefont {McLaughlin}}, \bibinfo {author} {\bibfnamefont {S.}~\bibnamefont {Li}}, \bibinfo {author} {\bibfnamefont {J.~A.}\ \bibnamefont {Brock}}, \bibinfo {author} {\bibfnamefont {S.}~\bibnamefont {Zhang}}, \bibinfo {author} {\bibfnamefont {H.}~\bibnamefont {Lu}}, \bibinfo {author} {\bibfnamefont {M.}~\bibnamefont {Huang}}, \bibinfo {author} {\bibfnamefont {Y.}~\bibnamefont {Xiao}}, \bibinfo {author} {\bibfnamefont {J.}~\bibnamefont {Zhou}}, \bibinfo {author} {\bibfnamefont {Y.}~\bibnamefont {Tserkovnyak}}, \bibinfo {author} {\bibfnamefont {E.~E.}\ \bibnamefont {Fullerton}}, \bibinfo {author} {\bibfnamefont {H.}~\bibnamefont {Wang}},\ and\ \bibinfo {author} {\bibfnamefont {C.~R.}\ \bibnamefont {Du}},\ }\bibfield  {title} {\bibinfo {title} {Local {{Control}} of a {{Single Nitrogen-Vacancy Center}} by {{Nanoscale Engineered Magnetic Domain Wall Motion}}},\ }\href {https://doi.org/10.1021/acsnano.3c10633} {\bibfield  {journal} {\bibinfo
  {journal} {ACS Nano}\ }\textbf {\bibinfo {volume} {17}},\ \bibinfo {pages} {25689} (\bibinfo {year} {2023})}\BibitemShut {NoStop}%
\bibitem [{\citenamefont {Kim}\ \emph {et~al.}(2025)\citenamefont {Kim}, \citenamefont {Rathi}, \citenamefont {Zhang}, \citenamefont {Seth}, \citenamefont {Simonov}, \citenamefont {Rutherford}, \citenamefont {Chen}, \citenamefont {Zhou}, \citenamefont {Peng}, \citenamefont {Xu}, \citenamefont {Xie}, \citenamefont {Vira}, \citenamefont {Tian}, \citenamefont {Ozerov}, \citenamefont {Kimchi}, \citenamefont {Mourigal}, \citenamefont {Smirnov},\ and\ \citenamefont {Jiang}}]{kim_sharp_2025}%
  \BibitemOpen
  \bibfield  {author} {\bibinfo {author} {\bibfnamefont {C.}~\bibnamefont {Kim}}, \bibinfo {author} {\bibfnamefont {S.}~\bibnamefont {Rathi}}, \bibinfo {author} {\bibfnamefont {N.}~\bibnamefont {Zhang}}, \bibinfo {author} {\bibfnamefont {A.}~\bibnamefont {Seth}}, \bibinfo {author} {\bibfnamefont {N.~V.}\ \bibnamefont {Simonov}}, \bibinfo {author} {\bibfnamefont {A.}~\bibnamefont {Rutherford}}, \bibinfo {author} {\bibfnamefont {L.}~\bibnamefont {Chen}}, \bibinfo {author} {\bibfnamefont {H.}~\bibnamefont {Zhou}}, \bibinfo {author} {\bibfnamefont {C.}~\bibnamefont {Peng}}, \bibinfo {author} {\bibfnamefont {M.}~\bibnamefont {Xu}}, \bibinfo {author} {\bibfnamefont {W.}~\bibnamefont {Xie}}, \bibinfo {author} {\bibfnamefont {A.~D.}\ \bibnamefont {Vira}}, \bibinfo {author} {\bibfnamefont {M.}~\bibnamefont {Tian}}, \bibinfo {author} {\bibfnamefont {M.}~\bibnamefont {Ozerov}}, \bibinfo {author} {\bibfnamefont {I.}~\bibnamefont {Kimchi}}, \bibinfo {author} {\bibfnamefont {M.}~\bibnamefont {Mourigal}}, \bibinfo {author}
  {\bibfnamefont {D.}~\bibnamefont {Smirnov}},\ and\ \bibinfo {author} {\bibfnamefont {Z.}~\bibnamefont {Jiang}},\ }\href@noop {} {\bibinfo {title} {Sharp spectroscopic fingerprints of disorder in an incompressible magnetic state}} (\bibinfo {year} {2025}),\ \Eprint {https://arxiv.org/abs/2506.08112} {arXiv:2506.08112 [cond-mat.mtrl-sci]} \BibitemShut {NoStop}%
\bibitem [{\citenamefont {Xu}\ \emph {et~al.}(2025)\citenamefont {Xu}, \citenamefont {Seth},\ and\ \citenamefont {Kimchi}}]{xu_data_2025}%
  \BibitemOpen
  \bibfield  {author} {\bibinfo {author} {\bibfnamefont {R.}~\bibnamefont {Xu}}, \bibinfo {author} {\bibfnamefont {A.}~\bibnamefont {Seth}},\ and\ \bibinfo {author} {\bibfnamefont {I.}~\bibnamefont {Kimchi}},\ }\href {https://doi.org/10.35090/gatech/79984} {\bibinfo {title} {Data for {{Chirality}} reversal at finite magnetic impurity strength and local signatures of a topological phase transition: {{https://doi.org/10.35090/gatech/79984}}}} (\bibinfo {year} {2025})\BibitemShut {NoStop}%
\end{thebibliography}%

\end{document}